\documentclass[fleqn,usenatbib]{mnras}

\usepackage{mathptmx}
\usepackage{relsize}
\usepackage{longtable}
\usepackage{cuted}
\usepackage{stfloats}
\usepackage{threeparttablex}
\usepackage{lipsum}

\usepackage[T1]{fontenc}
\usepackage{soul}

\DeclareRobustCommand{\VAN}[3]{#2}
\let\VANthebibliography\thebibliography
\def\thebibliography{\DeclareRobustCommand{\VAN}[3]{##3}\VANthebibliography}

\usepackage{graphicx}	
\usepackage{amsmath}	
\usepackage{amssymb}	
\usepackage{mathtools}
\usepackage{float}
\usepackage{graphicx}
\usepackage{ulem}
\usepackage{lscape}
\usepackage{xspace}

\title[Asynchronous accretion can mimic diverse white dwarf pollutants]{Asynchronous accretion can mimic diverse white dwarf pollutants I: core and mantle fragments}

\author[Brouwers Et Al.]{
	Marc G. Brouwers,$^{1}$\thanks{E-mail: mgb52@cam.ac.uk}
	Amy Bonsor$^{1}$, Uri Malamud$^{2,3}$
	\\
	$^{1}$Institute of Astronomy, University of Cambridge, Madingley Road, Cambridge CB3 0HA \\
	$^{2}$Department of Physics, Technion - Israel Institute of Technology, Technion City, 3200003 Haifa, Israel\\
	$^{3}$School of the Environment and Earth Sciences, Tel Aviv University, Ramat Aviv, 6997801 Tel Aviv, Israel
}

\date{Accepted 29 Oct 2022. Received 24 Oct 2022; in original form 18 Aug 2022}

\pubyear{2022}

\begin{document}
	\maketitle
	
	\begin{abstract}
        Polluted white dwarfs serve as astrophysical mass spectrometers – their photospheric abundances are used to infer the composition of planetary objects that accrete onto them. We show that due to asymmetries in the accretion process, the composition of the material falling onto a star may vary with time during the accretion of a single planetary body. Consequently, the instantaneous photospheric abundances of white dwarfs do not necessarily reflect the bulk composition of their pollutants, especially when their diffusion timescales are short. In particular, we predict that when an asteroid with an iron core tidally disrupts around a white dwarf, a larger share of its mantle is ejected, and that the core/mantle fraction of the accreting material varies with time during the event. Crucially, this implies that the core fraction of differentiated pollutants cannot be determined for white dwarfs with short diffusion timescales, which sample only brief episodes of longer accretion processes. The observed population of polluted white dwarfs backs up the proposed theory. More white dwarfs have accreted material with high Fe/Ca than low Fe/Ca relative to stellar abundance ratios, indicating the ejection of mantle material. Additionally, we find tentative evidence that the accretion rate of iron decreases more rapidly than that of magnesium or calcium, hinting at variability of the accreted composition. Further corroboration of the proposed theory will come from the up-coming analysis of large samples of young white dwarfs.
	\end{abstract}
	
	

\section{Introduction}\label{sect:introduction}
At least a quarter of all white dwarfs are polluted – they show spectroscopic signs of heavy, rock-forming elements (i.e., Mg, Si, Ca, and Fe) in their photospheres that otherwise contain only H or He \citep{Zuckerman2003, Zuckerman2010, Koester2014, Wilson2019}. 
The standard interpretation is that these stars are sampling the remains of old planetary systems \citep{Jura2003, Jura2014, Farihi2016a, Veras2021b}, whose planets and asteroid belts beyond a few AU survived the expansion of the star on the red and asymptotic giant branches \citep{Duncan1998, Villaver2009, Veras2011, Veras2012, veras2016a} and were subsequently scattered close to the white dwarf via a range of mechanisms \citep{Bonsor2011, Debes2012a, Antoniadou2016, Mustill2018, Maldonado2020, Veras2021a}. In some systems, this circumstellar material emits a detectable infrared excess \citep{Rocchetto2015, Farihi2016a, Wilson2019, Rogers2020}, and there are several instances where the existence of circumstellar material is implied by the detection of transits \citep{Vanderburg2015, Manser2019b, Vanderbosch2020, Vanderbosch2021, Guidry2021, Farihi2022, Budaj2022}. Recently, ongoing accretion onto a white dwarf was confirmed via the detection of X-rays \citep{Cunningham2022}.

Observations of polluted white dwarfs with multiple identified photospheric elements are of special significance because they provide the only direct compositional measurement of exoplanetary material. In most cases, the inferred pollutant abundances roughly resemble those of Earth \citep{Zuckerman2007, Klein2010, Xu2014b, Xu2019, Doyle2019}, although in some cases they hint at substantially different geological histories \citep{Putirka2021}. Notably, several systems contain increased levels of volatile elements (O, C, N), possibly indicating the accretion of cometary material that condensed in the outer regions of planetary discs \citep{Farihi2013, Raddi2015, Xu2017}. Other pollutants contain more refractory species (Ca, Ti) and likely formed on closer orbits \citep{Xu2014b}. In GD362, the particularly high ratio of Mn/Na could indicate a history of post-nebula volatilization \citep{Harrison2021a}, similar to the process experienced by Mars and the Moon \citep{Palme2003, Siebert2018}
\begin{figure*}
\centering
\includegraphics[width=\hsize]{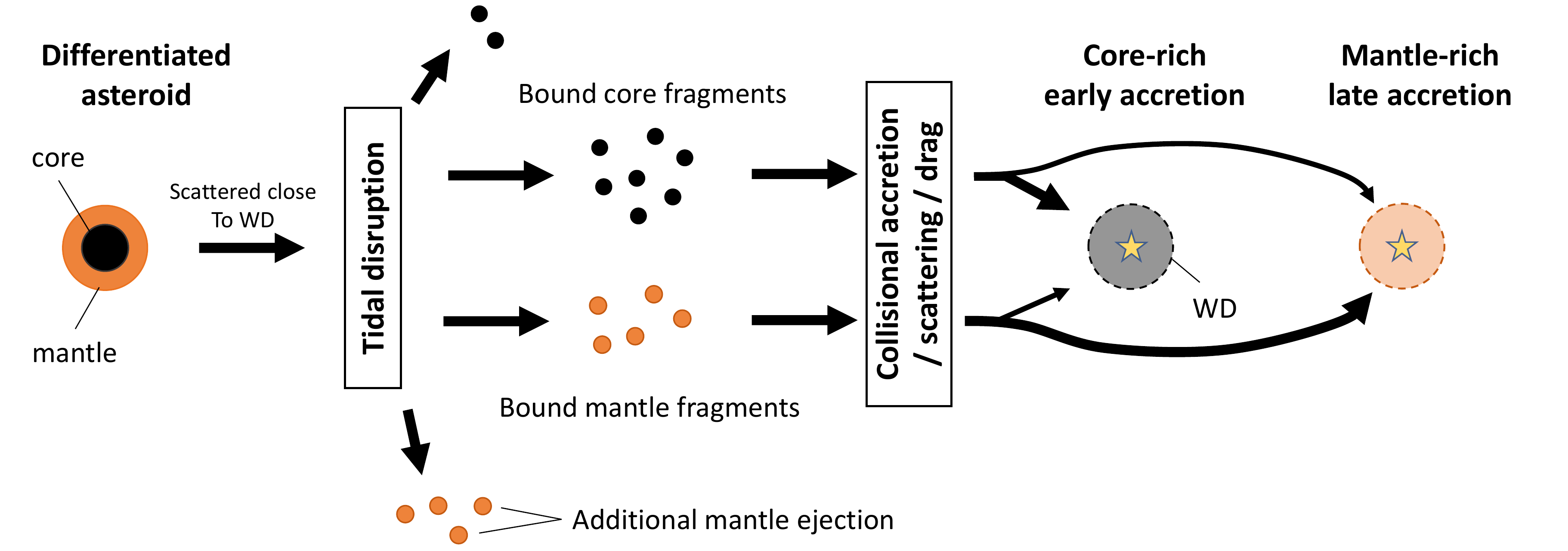} 
\caption{Schematic overview of how the asynchronous accretion of a differentiated asteroid onto a white dwarf can mimic the photospheric signature of either a core- or mantle-rich object. In this simplified scenario, a differentiated asteroid tidally disrupts into fragments composed of core or mantle material. Mantle fragments are preferentially ejected out of the system in greater proportions, and bound core and mantle fragments occupy distinct orbital zones in the tidal disc (Section \ref{sect:tidal_disruption}). Core fragments collide or scatter sooner on average, leading to a core-rich early accretion phase, followed by a mantle-rich late accretion phase (Section \ref{sect:accretion}). The form of this figure was inspired by the conceptually similar Fig. 1 of \citet{Buchan2022}. An accompanying version of this figure that details asynchronous ice-refractory accretion is presented in an accompanying paper II.
\label{fig:schematic}}
\end{figure*}

In addition to probing volatility trends, polluted white dwarfs can also reveal the proclivity of differentiation in exo-planetesimals \citep{Zuckerman2011, Jura2013, Bonsor2020}. Objects differentiate when their interiors become hot enough to partially melt, and chemical species separate into an iron-rich core and a magnesium-silicate-rich mantle. The decay of short-lived radioactive nuclei, such as ${}^{26}$Al, as witnessed in the Solar System, can fuel large-scale melting in asteroids larger than $\sim10$ km \citep{Hevey2006, Lichtenberg2016}. When a white dwarf's photosphere contains an over-abundance of iron and other siderophile (iron-loving – Ni, Cr) elements, it indicates that the star has swallowed a core-rich body \citep{Melis2011, Gaensicke2012, Wilson2015, Hollands2018, Hollands2021}. High abundances of lithophile (rock-loving – Ca, Mg, Si) elements instead hint at the accretion of predominantly mantle material, although this can be difficult to distinguish from a post-accretion (declining) phase \citep{Harrison2018, Harrison2021b, Buchan2022}. Pollutant abundances towards either extreme indicate that the parent body formed early or large enough to be differentiated and then experienced collisional processing. With the ever-increasing number of known polluted white dwarfs, statistics from larger samples of pollutants are expected to reveal general trends in the process of planet formation. 

In order to accurately translate spectroscopic signatures into pollutant abundances, a detailed understanding of photospheric physics is required, with well-constrained diffusion timescales for different elements \citep{Koester2014, Koester2020, Cunningham2021}. It is equally important, however, to understand how accretion onto white dwarfs proceeds temporally. If different parts of a pollutant (e.g., core/mantle, volatile/refractory) enter the white dwarf's photosphere at different times, this \textit{asynchronous accretion} could mimic an identical signature of e.g., a core-rich or a volatile-rich body whose parts accrete synchronously (see Fig. \ref{fig:schematic}). The implicit assumption in the current analyses of polluted white dwarfs – that accretion proceeds both synchronously and symmetrically – is clearly valid when a solid body directly strikes the white dwarf's surface \citep{Brown2017, Mcdonald2021} but such instances are predicted to be exceedingly rare \citep{Veras2021a}. Instead, the accretion process is thought to begin with a tidal disruption and the formation of an eccentric tidal disc \citep{Debes2012a, Veras2014, Nixon2020, Malamud2020a, Malamud2020b}, whose fragments can subsequently accrete via a range of processes \citep[e.g.,][]{Veras2015a, Veras2015b,Malamud2021,Li2021, Brouwers2022a}. In this work, we perform the first analysis of the synchronicity of accretion onto white dwarf photospheres. This first paper considers the accretion of core and mantle fragments from a differentiated pollutant. In an accompanying paper (\citet{Brouwers2022c}, henceforth paper II), we study the contrast in accretion between refractory materials and ices (e.g., $\mathrm{H_2O}, \mathrm{CO_2}$). Our results highlight that different elements are theoretically expected to accrete in proportions that vary over time, a finding that is supported by our analysis of the current sample of polluted whited dwarfs, and that can be corroborated further with upcoming large samples of young white dwarfs.

This paper is organized as follows. We first show in Section \ref{sect:tidal_disruption} how a tidal disruption unevenly spreads and ejects core and mantle fragments, providing an asymmetric starting point for the accretion process. We then consider the collisional grind-down of a differentiated asteroid in Section \ref{sect:grinddown} and evaluate the relative scattering of core and mantle fragments by a planet in Section \ref{sect:scattering}. In order to test the validity of our proposition, we investigate the accretion rate and abundance distribution of iron relative to lithophile elements in Section \ref{sect:observations}. We discuss our findings in Section \ref{sect:discussion} and conclude in Section \ref{sect:conclusions}.

\section{Disruption of differentiated asteroids}\label{sect:tidal_disruption}
\begin{figure*}
\centering
\includegraphics[width=\hsize]{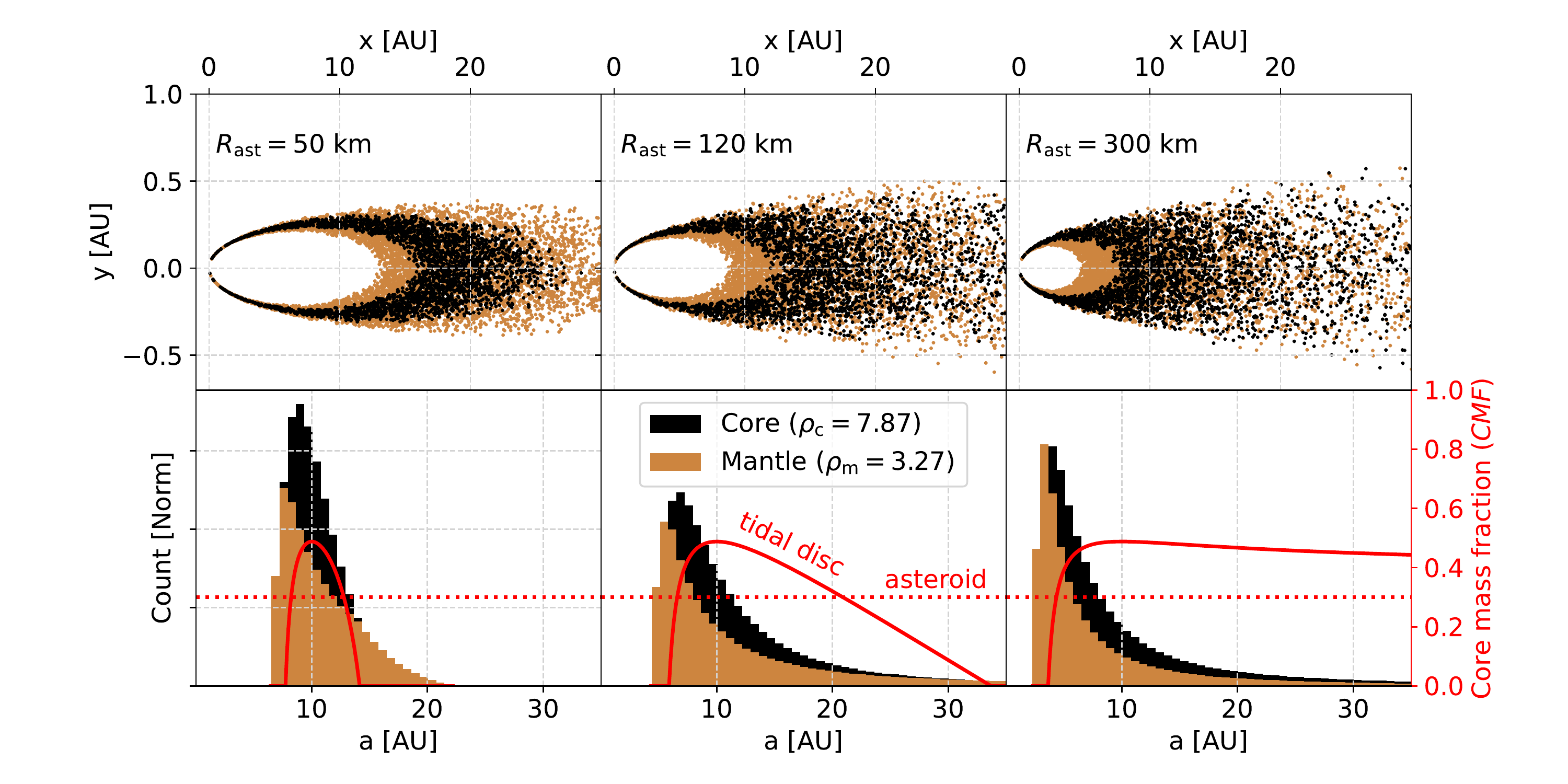} 
\caption{Distribution of core and mantle fragments after the tidal disruption of a differentiated asteroid from 10 AU around a white dwarf. The top panels provide a top-down view of the tidal disc, and show that core (black) and mantle (brown) fragments are spread to distinct, but partially overlapping orbital ranges. The histogram plots in the lower panels show the distribution of material in more detail. The solid red line indicates the core mass fraction ($CMF$) at a given orbit, which can be compared to the 30\% CMF of the asteroid progenitor (red dotted line). The innermost orbits of the tidal disc always contain exclusively mantle material, while the outer orbits are enhanced in core material if the asteroid was large (right panels), and depleted otherwise (left and middle panels). \label{fig:disruption_geometries}}
\end{figure*}
The accretion process of planetary material onto a white dwarf is thought to begin with the perturbation of an asteroid or planetary body onto a highly eccentric orbit \citep[e.g.,][]{Bonsor2011, Mustill2018, Smallwood2018}. When the asteroid ventures too close to the star, its internal strength and self-gravity are overwhelmed by stellar gravity, and it tidally disrupts \citep{Debes2012a, Veras2014, Malamud2020a, Malamud2020b}. In this section, we consider a simplified scenario where the disrupted body was differentiated into two radially separate layers: a central core, and an outer mantle. We first study how these components are geometrically spread in the tidal disc, and then trace their accretion onto the white dwarf as a function of time. If both components (core and mantle) accrete with a constant mass ratio over time, the composition of the photosphere remains unchanged. However, if the accretion of either component follows a different trend, and the accretion process is \textit{asynchronous}, the composition of the photosphere varies with time during the accretion of a single body.

\subsection{Geometry of core and mantle fragments in tidal discs}
In the scenario that asteroids are torn apart by strong tidal forces close to the star, accretion is preceded by the formation of an eccentric tidal disc. Core and mantle fragments will occupy distinct orbits in this disc due to their different radial positions
within the asteroid. In order to trace the distribution of these orbits, we consider a simplified, instantaneous tidal disruption at a distance $r_\mathrm{B} = 0.75 \; \mathrm{R_\odot}$ from the star\footnote{Corresponding to the disruption of a strengthless or sufficiently large asteroid with a core mass fraction ($CMF$) of 0.3 around a 0.6 $M_\odot$ white dwarf \citep[e.g.,][]{Davidsson1999,Bear2013}.}. The energies of the fragments $i$ depend on their distance $r_\mathrm{i}$ to the star, and their semi-major axes $a_\mathrm{i}$ become spread out along the range \citep[e.g.,][]{Brouwers2022a}:
\begin{equation}
    a_\mathrm{i} = a_\mathrm{ast} {\left(1+ 2a_\mathrm{ast}\frac{r_\mathrm{B}- r_\mathrm{i}}{r_\mathrm{B}r_\mathrm{i}}\right)}^{-1}, \label{eq:a_new_2}
\end{equation}
with eccentricities $e_\mathrm{i} = 1 - r_\mathrm{i} / a_\mathrm{i}$. This calculation safely ignores the spin of the asteroid (see Appendix \ref{appendix:sg_rot}). While the fragments initially form in a cluster, with core and mantle regions not necessarily disrupting simultaneously \citep{Veras2017, Duvvuri2020, Malamud2020a}, they shear out over time, and completely fill a tidal disc after a well-defined timescale (see Appendix \ref{appendix:spreading} for a derivation of this filling time). The width of the tidal disc that forms in this manner depends chiefly on the size of the asteroid and on its semi-major axis. The fragments of small asteroids that originate from a planetary system spread out along a narrow orbital band when they disrupt \citep{Veras2014,Veras2021a,Nixon2020}, while larger objects on wider orbits form a broader tidal disc, culminating in a completely bimodal disruption for planet-sized bodies, where half of their fragments eject from the system, and the rest become concentrated close to the star \citep{Rafikov2018, Malamud2020a, Malamud2020b}.

Fragments from different layers in the asteroid are spread to distinct orbits. Therefore, an asteroid that has any radial variation in its composition will form a tidal disc whose geometry retains a similar compositional variation. We illustrate this asymmetry in Fig. \ref{fig:disruption_geometries} for a range of asteroid sizes, assuming representative core and mantle densities of $\rho_\mathrm{c}=7.87 \;\mathrm{g/cm^3}$ and $\rho_\mathrm{m}=3.27 \;\mathrm{g/cm^3}$, respectively, corresponding to iron and forsterite. Small asteroids (left panel, $R_\mathrm{ast} = 50$ km), disrupt into a tidal disc whose intermediate orbits contain additional core material, whereas both its inner and outer orbits consist entirely of mantle fragments. Larger asteroids of 120 km (middle panel) form tidal discs whose core fragments spread out over a wider range, although their innermost and outermost fragments still consist exclusively of mantle material. At $R_\mathrm{ast} = 300$ km (right panel), a substantial fraction of the fragments become unbound from the stellar system and all except the closest orbits are enhanced in core material.

\subsection{Ejection bias of mantle and crustal fragments}\label{subsect:ejection}
The parts of the asteroid that are furthest from the white dwarf during the tidal disruption are the easiest to eject. The outer layers of a differentiated asteroid are part of its mantle, and so the remaining material that forms the bound tidal disc can become dominated by the core. The dividing line between bound and unbound fragments is drawn at a distance $R_\mathrm{eject}$ from the asteroid's centre \citep{Malamud2020a}:
\begin{equation} \label{eq:R_eject}
    R_\mathrm{eject} = \frac{r_\mathrm{B}^2}{2a_\mathrm{ast}-r_\mathrm{B}},
\end{equation}
with all the material beyond $R_\mathrm{eject}$ lost into space. To compute the core mass fraction of the bound tidal disc, we compare the total volume fraction of the ejected material ($\chi_\mathrm{ast} = V_\mathrm{ast, unbound} / V_\mathrm{ast}$) to the ejection fraction of the core ($\chi_\mathrm{c} = V_\mathrm{c, unbound} / V_\mathrm{c}$):
\begin{subequations}
\begin{align}
    \chi_\mathrm{ast} &= \frac{3}{4\pi R_\mathrm{ast}^3}
    \int_{R_\mathrm{ast,eject}}^{R_\mathrm{ast}} \pi \left(R_\mathrm{ast}^2-x^2\right) dx \\
    &= \frac{\left(R_\mathrm{ast}-R_\mathrm{ast,eject}\right)^2 \left(2 R_\mathrm{ast}+R_\mathrm{ast,eject}\right)}{4 R_\mathrm{ast}^3}, \\
    \chi_\mathrm{c} &= \frac{3}{4\pi R_\mathrm{c}^3}
    \int_{R_\mathrm{c,eject}}^{R_\mathrm{c}} \pi \left(R_\mathrm{c}^2-x^2\right) dx \\
    &= \frac{\left(R_\mathrm{c} - R_\mathrm{c,eject}\right)^2 \left(2 R_\mathrm{c} + R_\mathrm{c,eject}\right)}{4 R_\mathrm{c}^3},
\end{align}
\end{subequations}
where $R_\mathrm{c,eject} = \mathrm{min(}R_\mathrm{eject},R_\mathrm{c,ast})$ and $R_\mathrm{ast,eject} = \mathrm{min(}R_\mathrm{eject},R_\mathrm{ast})$. Combined, these expressions analytically specify the mass fraction of core material in the bound tidal disc ($CMF_\mathrm{disc}$):
\begin{subequations}
\begin{align}
    CMF_\mathrm{disc} &= \frac{\rho_\mathrm{c}\left(V_\mathrm{c}-V_\mathrm{c,unbound}\right)}{\left(\rho_\mathrm{c} - \rho_\mathrm{m}\right) \left(V_\mathrm{c}-V_\mathrm{c,unbound}\right) + \rho_\mathrm{m}\left(V_\mathrm{ast}-V_\mathrm{ast,unbound}\right)} \\
    &= \frac{1-\chi_\mathrm{c}}{\left(\chi_\mathrm{ast}-\chi_\mathrm{c}\right) \left(1-\rho_\mathrm{m}/\rho_\mathrm{c} \right) + \left(1-\chi_\mathrm{ast}\right)/CMF_\mathrm{ast}}, \label{eq:CMF_bound}
    \end{align}
\end{subequations}
where $CMF_\mathrm{ast} = M_\mathrm{c,ast} / M_\mathrm{ast}$ is the core mass fraction of the asteroid.
\begin{figure}
\centering
\includegraphics[width=\hsize]{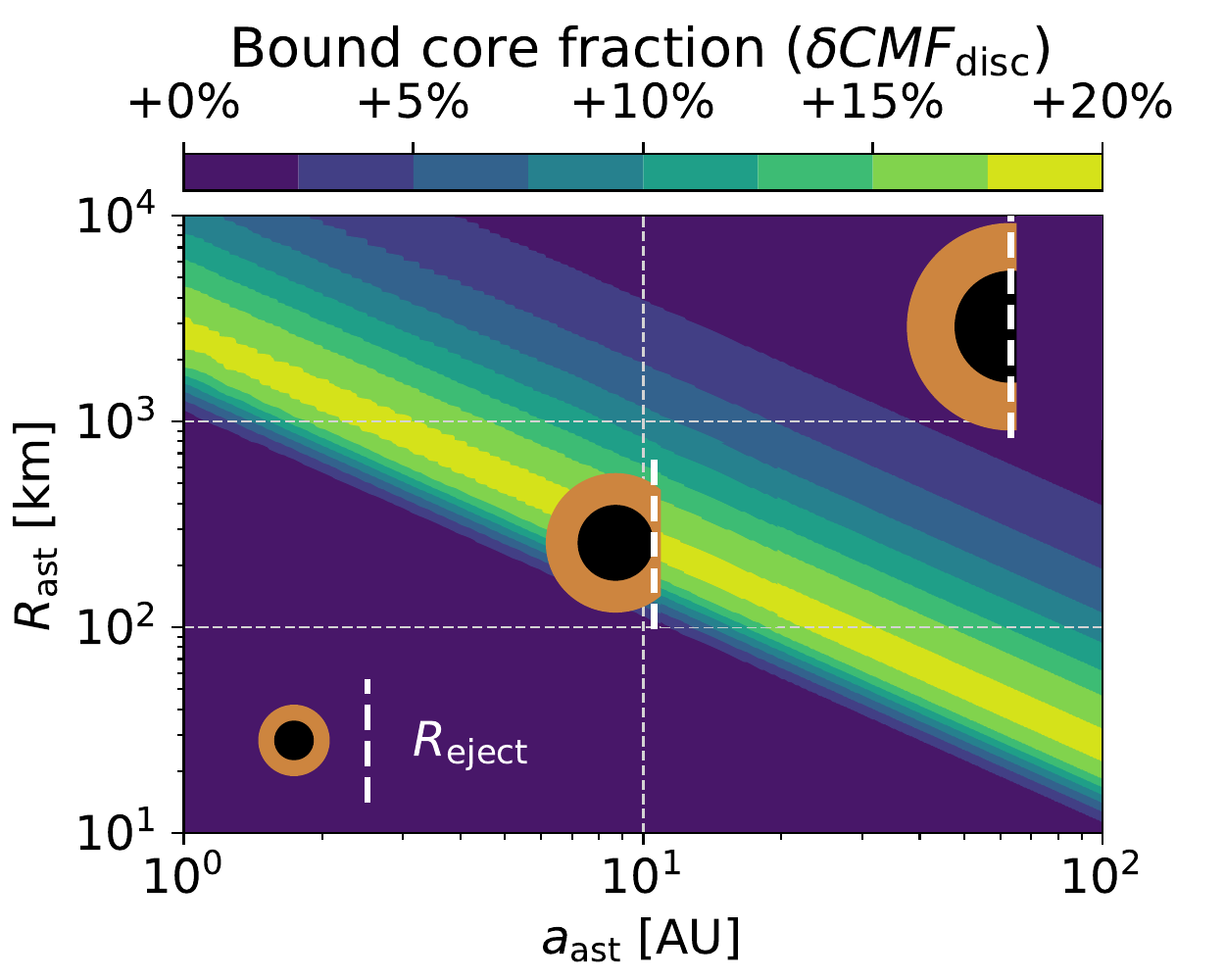}
\caption{Enhancement in bound core mass fraction ($\delta CMF_\mathrm{disc} = \left(CMF_\mathrm{disc} - CMF_\mathrm{ast}\right)/CMF_\mathrm{ast}$, colours) of tidal discs compared to their asteroid progenitors. Three regimes can be identified. Both sufficiently small asteroids and large planetary bodies disrupt into discs with unaltered core mass fractions, as either half or none of their mass is ejected. In between, larger asteroids and dwarf planets eject primarily crustal and mantle material when they disrupt, resulting in a more core-rich tidal disc. \label{fig:core_increase_BulkEarth}}
\end{figure}

We show the core enhancement of the tidal disc in Fig. \ref{fig:core_increase_BulkEarth}, which presents three distinct regimes as a function of asteroid size and semi-major axis. Fragments from sufficiently small or tight asteroids ($R_\mathrm{ast} < R_\mathrm{eject}$) remain entirely bound to the star, and the core mass fraction of their tidal discs is equal to that of their asteroid progenitors. Similarly, very large asteroids ($R_\mathrm{ast} \gg R_\mathrm{eject}$) disrupt in a bimodal manner where nearly half of their material is ejected, again leaving the bulk core fractions in the tidal disc largely unchanged. In between, however, intermediate-sized asteroids ($R_\mathrm{ast} \sim R_\mathrm{eject}$) eject almost exclusively mantle and crustal fragments, increasing the fraction of core material in their tidal disc by up to 20\%. The size range most affected by this ejection asymmetry lies between 100-1000 km in the inner disc ($<10$ AU), which overlaps with the larger asteroids and dwarf planets in the Solar System's asteroid belt, like Vesta and Ceres. In the outer disc, the affected sizes shrink substantially to $\sim20$ km at 100 AU. At this point, only the asteroids that formed sufficiently early will have differentiated, and so the importance of the ejection asymmetry gradually diminishes towards the far outer disc. Finally, we note that the ejection asymmetry, illustrated with core-mantle differentiated bodies, applies to any pollutant with a radial variation in composition. This includes the ejection of an increased portion of crustal material relative to both the core and mantle, as well as the ejection of additional ice when a comet contains an icy outer layer (see paper II). Therefore, when a differentiated object disrupts around a white dwarf, the material that accretes onto it will often not exactly match its bulk composition.

\section{Asynchronous accretion of core and mantle fragments}\label{sect:accretion}
In this section, we illustrate how the spatial asymmetry between core and mantle fragments in a tidal disc can cause these components to accrete asynchronously onto the white dwarf. We follow the road-map to accretion outlined by \citet{Brouwers2022a}, where accretion either proceeds via differential precession and collisional grind-down, or via the scattering of fragments by a planet (see also \citealt{Li2021}). In Section \ref{sect:alternative_models}, we discuss how core-mantle accretion may play out in different accretion models.

\subsection{Scenario I: differential precession and collisional grind-down}\label{sect:grinddown}
In this first scenario, we consider the three-stage accretion model suggested by \citet{Brouwers2022a}. In the first stage, a tidal disruption spreads the fragments over a range of highly eccentric orbits, as discussed in the previous section. The orbits do not follow precise Keplerian tracks and their pericentres precess over time due to GR, at rates that depend on their semi-major axis and eccentricity \citep{Debes2012a, Veras2014}. Inner fragments precess more quickly than those on wider orbits and unless interactions allow the disc to precess coherently, significant apsidal differences between fragments build up over time, causing orbits to cross. In the second stage of the model, fragments grind into dust at the intersection points. Finally, the dust quickly accretes due to drag forces, preventing a large infrared excess. In our calculation, we divide the fragments into a two-dimensional grid along semi-major axis and fragment size. The semi-major axis grid points accommodate a constant portion of fragment mass when the tidal disc forms. The fragment orbits are set by Eq. \ref{eq:a_new_2}, which we evaluate according to the spatial distribution of core and mantle material in a spherical asteroid. This is a significant improvement over the equal-energy fragment distribution assumed by \citealt{Brouwers2022a}. Orbits whose fragments originate from the centre of the asteroid contain more core material, whereas orbits contain additional mantle material (See Fig. \ref{fig:disruption_geometries}). The rate of catastrophic collisions is modelled with a particle-in-cell approach (Eqs. 22-26 of \citealt{Brouwers2022a}), with the crude assumption that a catastrophic collision turns the entire fragment into dust, which then quickly accretes onto the star.
\begin{figure}
\centering
\includegraphics[width=\hsize]{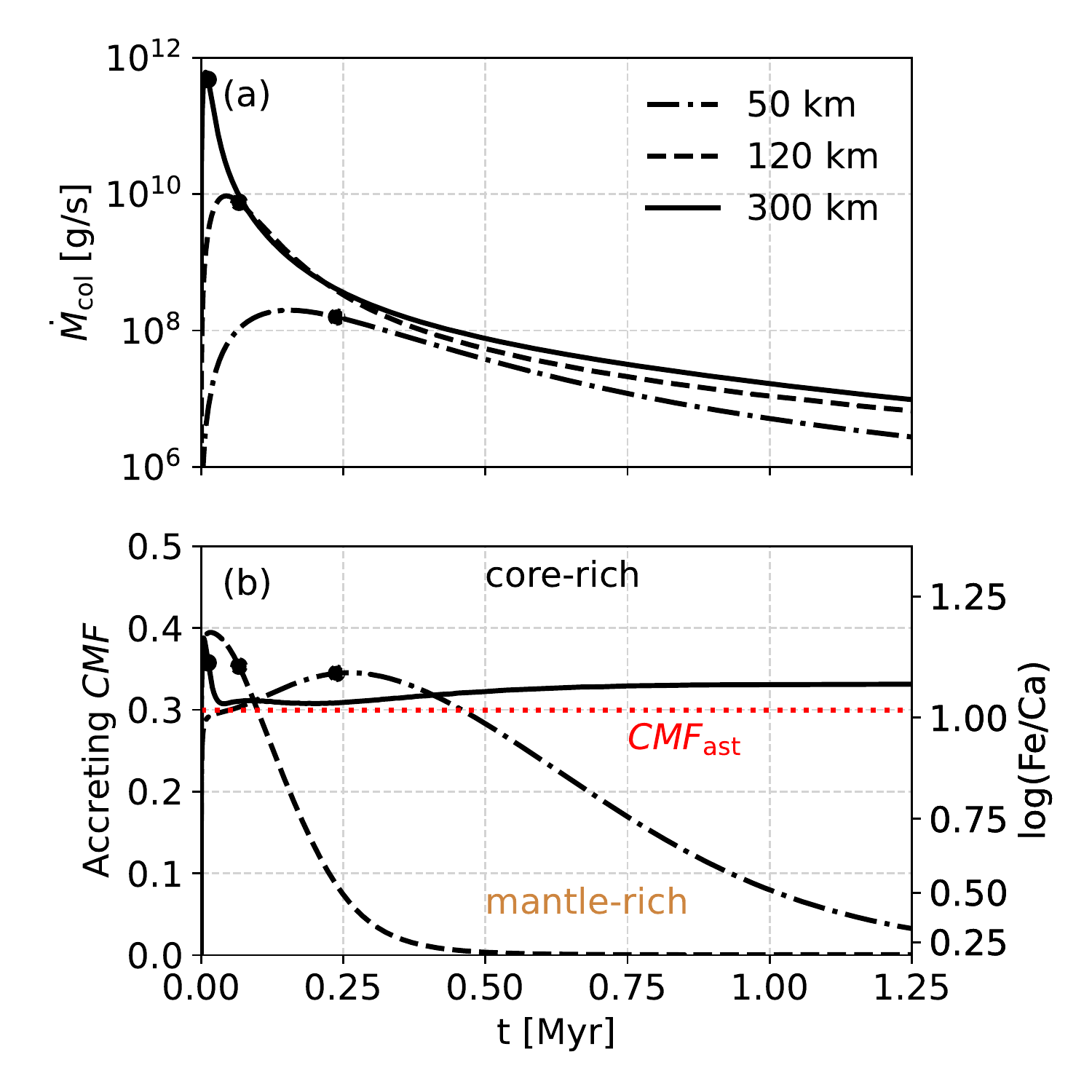}
\caption{The rate at which mass is lost due to collisions from the tidal disc, a proxy for the accretion rate onto polluted white dwarfs, calculated according to scenario I. The lines correspond to either a differentiated 50 km (dotted), 120 km (dashed) or 300 km (solid) asteroid, with semi-major axes at 10 AU and a 30\% core mass fraction (red line). The points indicate when half of the pollutant's mass has been accreted. In panel (a), the total collision rate first increases as fragment orbits begin to cross, and then decreases when the disc becomes depleted. In panel (b), the core mass fraction of the colliding material is shown to vary over time, leading to the asynchronous accretion of core and mantle components.\label{fig:accretion_curves}}
\end{figure}

We show the evolution of the total catastrophic collision rate of core and mantle fragments combined in panel (a) of Fig. \ref{fig:accretion_curves}, plotted for disrupted asteroids of 50, 120 and 300 km in size. These simulations are ran with 200 semi-major axis bins and 150 size bins, for a total of resolution of 30,000 fragments. As was shown in \citet{Brouwers2022a}, the collision rate first builds up, due to an increase in orbit crossings, and then declines when the tidal disc becomes depleted of fragments. The peak collision rate is much greater in more massive discs, as larger asteroids break up into more fragments that each collide more easily. In panel (b), we plot the core mass fraction of the accreting material as a function of time, with the equivalent Ca/Fe number ratio that would be observed in the photosphere if the core and mantle have Earth-like elemental abundances\footnote{Assuming steady-state accretion with a typical sinking timescale ratio between Ca and Fe of $\tau_\mathrm{Ca}/\tau_\mathrm{Fe}=1.4$ \citep{Koester2020}. Earth's core mean molecular weight is taken as $\mu_\mathrm{c}= 53.79$, with core Ca and Fe mass fractions of $f_\mathrm{Ca,c} = 0\%$ and $f_\mathrm{Fe,c} = 88.8\%$ \citep{Workman2005}. Earth's mantle equivalents are taken as: $\mu_\mathrm{m}=51.7$, $f_\mathrm{Ca,m} = 2.3\%$, and $f_\mathrm{Fe,m} = 6.4\%$ \citep{Morgan1980}.}. Core fragments in a tidal disc cluster on intermediate orbits, while those from the mantle mostly occupy the inner and outer orbits (see Fig. \ref{fig:disruption_geometries}). As a result, collisions initially involve an increased fraction of core fragments. The fragments located on the outermost orbits can spend many years at their apocentres every orbit, and are the last to collide. In the 50 km and 120 km examples, these outermost orbits only contain mantle fragments, and the core fraction of their accreting material drops all the way to zero over time. This is not the case for the grind-down of a larger 300 km asteroid (see Section \ref{sect:tidal_disruption}). 

The instantaneous accretion rates in Fig. \ref{fig:accretion_curves} indicate what would be observed in white dwarfs with short diffusion timescales, such as those with hydrogen-dominated atmospheres. The implied accretion rates and compositions will show reduced variability for older stars with helium-dominated atmospheres, as material remains in their atmospheres for longer, averaging over a portion of the accretion curve. When the diffusion timescale exceeds 1 Myr in this model, as is the case for helium-dominated envelopes cooler than 20,000 K, the entire accretion is averaged over, to a value closer to the asteroid's bulk composition.

\subsection{Scenario II: scattering of core fragments by a planet}\label{sect:scattering}
If a white dwarf is polluted by a planetesimal that was scattered onto a star-grazing orbit by a planet, it is likely that the planet continues to scatter the fragments after the main body is disrupted. This alternative accretion scenario was discussed by \citet{Brouwers2022a} and studied with more numerical simulations by \citet{Li2021}. When a fragment is scattered by a planet, it can either collide with the white dwarf, be ejected from the system, or just continue on a different orbit \citep{Wyatt2017}. In the scattering simulations by \citet{Li2021} with a Neptune-mass planet, most scattered fragments quickly hit the white dwarf. We suggest that core fragments are significantly more likely to be re-scattered by a planet due to their central positioning in the tidal disc, providing a second channel for asynchronous core-mantle accretion.

In the example that we study here, a single planet of 10 $\mathrm{M_\oplus}$ is located at the apocentre of the asteroid's orbit ($a_\mathrm{pl} = 2 a_\mathrm{ast}$). The maximum allowed distance where fragments are re-scattered can be approximated from the width of the chaotic zone ($\delta a_\mathrm{chaos} = C \; a_\mathrm{pl} \left(\frac{M_\mathrm{pl}}{M_\mathrm{WD}}\right)^\frac{2}{7}$ \citep{Wisdom1980, Duncan1989, Quillen2006, Chiang2009}), with constant $1.3 < C < 2$. The inner edge of the chaotic zone provides a minimum semi-major axis for re-scattering $a_\mathrm{i}>a_\mathrm{cross}$ \citep{Brouwers2022a}, with:
\begin{subequations}
\begin{equation}
    a_\mathrm{cross} \geq \frac{a_\mathrm{pl}}{2}\left[1-C \left(\frac{M_\mathrm{pl}}{M_\mathrm{WD}}\right)^\frac{2}{7}\right].
\end{equation}
\end{subequations}
The semi-major axis of a fragment is related to its position relative to the star at the moment of disruption via Eq. \ref{eq:a_new_2}. Using that $r_\mathrm{B} \ll a_\mathrm{ast}$, we can derive an analogous criterion to the ejection distance (Eq. \ref{eq:R_eject}) for the re-scattering of fragments by a planet. If the distance of a fragment to the white dwarf exceeds $r_\mathrm{B} + R_\mathrm{scat}$ at the moment of the tidal disruption, the fragment will cross the chaotic zone of the planet:
\begin{subequations}
\begin{equation}
    R_\mathrm{scat} = -\frac{r_\mathrm{B}^2 \delta a_\mathrm{chaos}}{a_\mathrm{pl} (a_\mathrm{pl}-\delta a_\mathrm{chaos})}.
\end{equation}
\end{subequations}
These fragments that intersect a planet's orbit are susceptible to re-scattering. Similar to the calculation presented in Section \ref{subsect:ejection}, we can use this characteristic distance to compute the total volume fraction of fragments that can be scattered ($\widetilde{\chi}_\mathrm{ast} = V_\mathrm{tot, scat} / V_\mathrm{ast}$), and compare this to the fraction of core material liable to scattering ($\widetilde{\chi}_\mathrm{c} = V_\mathrm{c, scat} / V_\mathrm{c}$):
\begin{subequations}
\begin{align}
    \widetilde{\chi}_\mathrm{ast} &= \frac{3}{4\pi R_\mathrm{ast}^3}
    \int_{R_\mathrm{ast,scat}}^{R_\mathrm{ast,eject}} \pi \left(R_\mathrm{ast}^2-x^2\right) dx \\
    &= \frac{3 R_\mathrm{ast}^2\left(R_\mathrm{ast,eject}-R_\mathrm{ast,scat} \right) - R_\mathrm{ast,eject}^3+R_\mathrm{ast,scat}^3}{4 R_\mathrm{ast}^3}, \\
    \widetilde{\chi}_\mathrm{c} &= \frac{3}{4\pi R_\mathrm{c}^3}
    \int_{R_\mathrm{c,scat}}^{R_\mathrm{c,eject}} \pi \left(R_\mathrm{c}^2-x^2\right) dx \\
    &= \frac{3 R_\mathrm{c}^2\left(R_\mathrm{c,eject}-R_\mathrm{c,scat} \right) - R_\mathrm{c,eject}^3+R_\mathrm{c,scat}^3}{4 R_\mathrm{c}^3},
\end{align}
\end{subequations}
\begin{figure}
\centering
\includegraphics[width=\hsize]{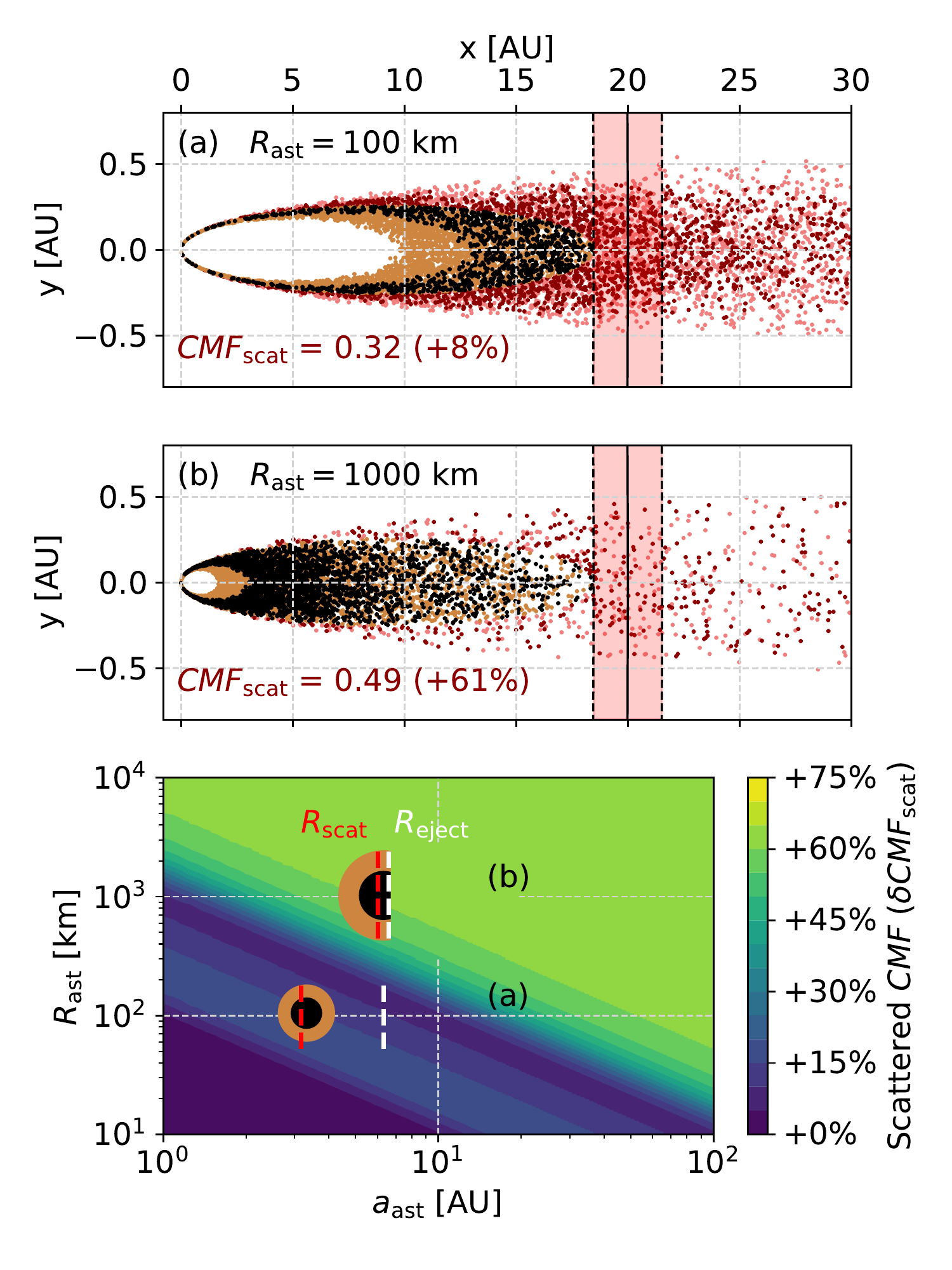}
\caption{Scattering asymmetry of core and mantle fragments by a 10 $\mathrm{M_\oplus}$ planet. In the top panels (a,b), the orbit of the planet and its chaotic zone are indicated by the red band. The red particles are susceptible to getting scattered, while brown (mantle) and black (core) particles are safe. On average, core fragments are more likely to be scattered by a planet, and the difference is greatest for larger asteroids and dwarf planets from the outer disc. \label{fig:scattering}}
\end{figure}
where $R_\mathrm{c,scat} = \mathrm{max(}R_\mathrm{scat},-R_\mathrm{c})$ and $R_\mathrm{ast,scat} = \mathrm{max(}R_\mathrm{scat},-R_\mathrm{ast})$. Together, these expressions specify the mass fraction of core material that can be re-scattered ($CMF_\mathrm{scat}$) from the tidal disc by a planet:
\begin{subequations}
\begin{align}
    CMF_\mathrm{scat} &= \frac{\rho_\mathrm{c} V_\mathrm{c,scat}}{\left(\rho_\mathrm{c}-\rho_\mathrm{m}\right) V_\mathrm{c,scat} + \rho_\mathrm{m} V_\mathrm{ast,scat}} \\
    &= \frac{\widetilde{\chi}_\mathrm{c}}{\left(\widetilde{\chi}_\mathrm{c}-\widetilde{\chi}_\mathrm{ast}\right) \left(1-\rho_\mathrm{m}/\rho_\mathrm{c}\right) + \widetilde{\chi}_\mathrm{ast}/CMF}.
\end{align}
\end{subequations}
We present the results of this calculation in Fig. \ref{fig:scattering}. Panels (a) and (b) show the disruption of a 100 km and 1000 km asteroid, respectively. In both cases, core fragments are more susceptible to re-scattering via close encounters with the planet, with an additional scattered core mass fraction between 8\% and 61\% present in the relevant orbital range. Fragments that are sufficiently tightly bound to the star are protected from close encounters with the planet, while unbound fragments eject on hyperbolic orbits. The fragments that remain to be scattered originate from the core-rich middle of the asteroid. While the scattering asymmetry in terms of core/mantle ratio is greatest for the largest bodies, only a small percentage of the bound fragments are scattered if the body is too large. Therefore, like the ejection asymmetry discussed in Section \ref{subsect:ejection}, the scattering asymmetry is most important for asteroids and dwarf planets in the range between 100-1000 km in the inner disc, and down to 20 km at 100 AU.

\section{Observational tests of asynchronous accretion}\label{sect:observations}
From the preceding arguments, the asymmetries in the accretion process of differentiated bodies onto white dwarfs lead to two observational predictions:
\begin{enumerate}
    \item Mantle fragments are preferentially ejected in the tidal disruption that likely precedes accretion. This implies that the material that accretes onto white dwarfs from differentiated bodies is enriched in core material by up to 20\%.
    \item Core and mantle fragments spread to distinct orbital ranges after a tidal disruption, causing them to accrete in a proportion that varies over time. Core fragments likely accrete faster on average than mantle fragments, so the core mass fraction of the accreted material is expected to decline over time during a single accretion event.
\end{enumerate}
In this section, we analyse with the current white dwarf sample to compare with these predictions.

\begin{figure}
\centering
\includegraphics[width=\hsize]{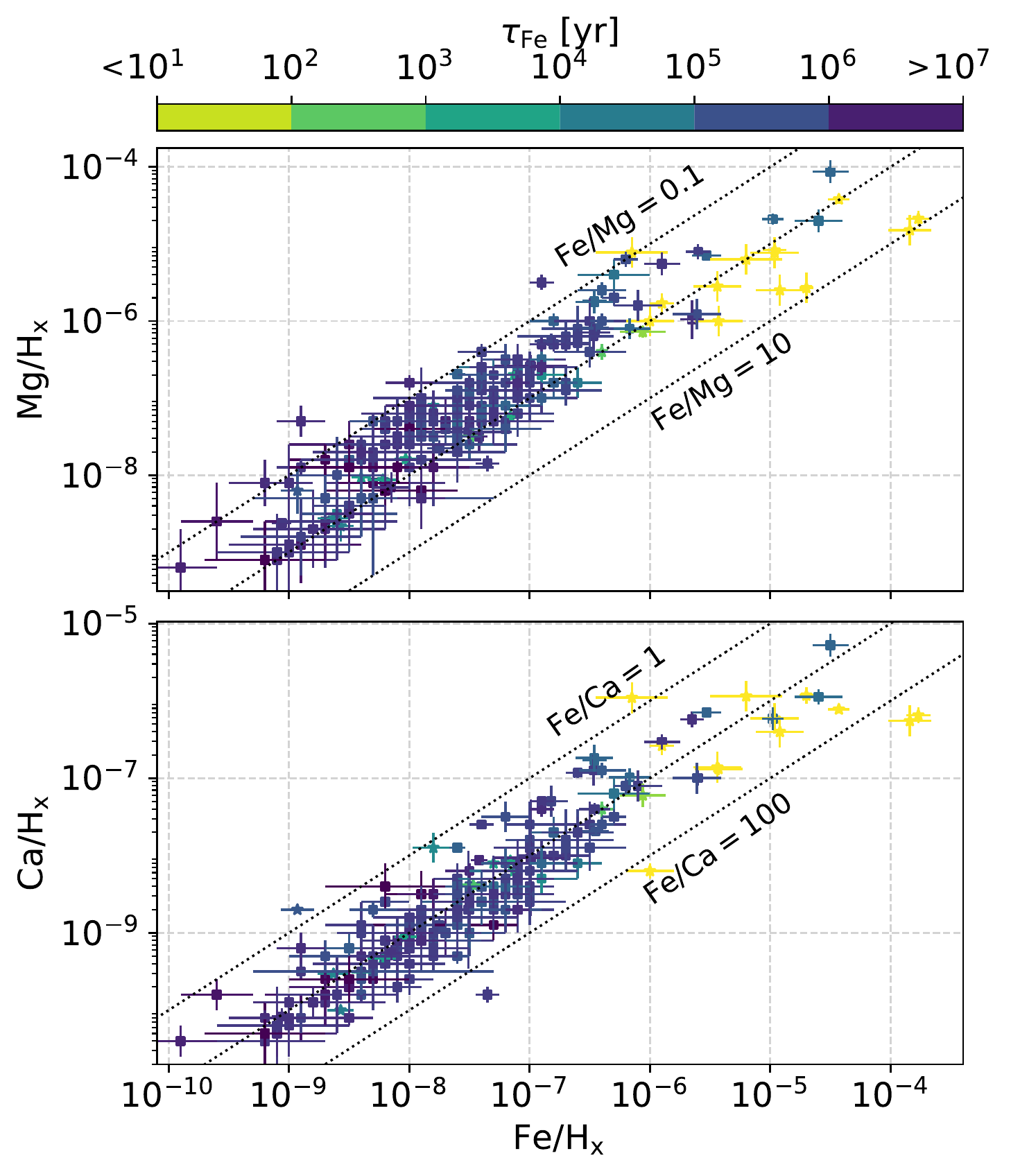}
\caption{Photospheric abundances of magnesium, calcium, and iron in our white dwarf sample. The colours indicate the diffusion timescale of iron in the white dwarf atmospheres, computed with models by \citet{Koester2020}. Lines of constant abundance ratios relative to iron are overplotted. \label{fig:CaFe_scatter}}
\end{figure}

\subsection{White dwarf sample with Fe/Mg or Fe/Ca abundance ratios}
To observationally study the accretion asymmetries between the core and mantle, we compile a sample of polluted white dwarfs with measured photospheric abundances of iron, in addition to Mg, and/or Ca. The sample is by no means uniformly selected, and contains all white dwarfs with these abundance ratios that we could find in the literature. Special care was taken to include white dwarfs with hydrogen-dominated atmospheres, so that the sample includes stars with a wide range of diffusion timescales, which vary between a day and several Myr in this sample. The photospheric abundances of the sample are plotted in Fig. \ref{fig:CaFe_scatter}, and show about two orders of magnitude variation in the ratios of Fe/Mg and Fe/Ca.

\subsection{Evidence for a core bias in the total accreted material}\label{sect:evidence_corebias}
We first study the first prediction, that the preferential ejection of mantle fragments after a tidal disruption (Section \ref{subsect:ejection}) skews the accretion of material onto white dwarfs to core-rich compositions. As a test, we compare the accretion rate ratios $\dot{M}_\mathrm{Fe}/\dot{M}_\mathrm{Ca}$ in the sample of polluted white dwarfs to a second sample of 957 nearby FGK stars \citep{Brewer2016, Harrison2018}, which represent a pristine abundance ratio. Because planetary material forms from the same molecular clouds that form stars, stellar data are a useful proxy for certain pristine abundance
ratios in a proto-planetary disc that are not altered by nucleosynthesis in the stars or incomplete condensation in the proto-planetary disc \citep{Lodders2003, Adibekyan2021}. In our analysis, we limit the sub-sample studied to older white dwarfs, whose diffusion times are sufficiently long (>Myr) to average over the accretion process, allowing the bulk composition of the accreted material to be measured. Furthermore, we use Ca rather than Mg here because of significant differences in reported Mg diffusion times in the atmospheres of cool, helium-dominated white dwarfs \citep{Hollands2017, Turner2020, Blouin2020}.

In an equilibrium between accretion and downward diffusion, the accretion rate of element El onto a white dwarf is given by:
\begin{equation}\label{eq:Mdot_ss}
    \dot{M}_\mathrm{El} = \mathrm{El}/\mathrm{H_x} \frac{\mu_\mathrm{El} M_\mathrm{cvz}}{\mu_\mathrm{H_x}\tau_\mathrm{El}},
\end{equation}
where $\mathrm{El}/\mathrm{H_x}$ is the photospheric abundance of El relative to hydrogen or helium, $M_\mathrm{cvz}$ is the mass of the star's convective zone, $\mu_\mathrm{El}, \mu_\mathrm{H_x}$ are atomic weights and $\tau_\mathrm{El}$ is the diffusion timescale of El. In our sample, the abundances $\mathrm{El}/\mathrm{H_x}$ are collated from the literature (see Table \ref{table:sample}), and $M_\mathrm{cvz}, \tau_\mathrm{El}$ are calculated with the module \texttt{timescale\_interpolator} from the open source code \texttt{PyllutedWD}\footnote{\url{https://github.com/andrewmbuchan4/PyllutedWD_Public}} \citep{Harrison2021b, Buchan2022}, based on updated white dwarf models from \citet{Koester2020}. The photospheric abundances of Fe and Ca in white dwarfs are subject to significant errors. Assuming that these abundance errors are independent, and that the diffusion timescales are exactly known, the combined error on the accretion ratio is:
\begin{equation}\label{eq:sigma}
    \sigma_\mathrm{{log(\dot{M}_\mathrm{Fe}/\dot{M}_\mathrm{Ca})} = \sqrt{\sigma_{log(Fe/H_{x})}^2 + \sigma_{log(Ca/H_{x})}^2}}.
\end{equation}

We follow a slightly modified Kolmogorov–Smirnov test procedure to compare the abundance ratios of white dwarfs to pristine material. Assuming that the reported abundance errors represent normal distributions, the probability density $\mathrm{PDF}^\star$ of an accretion rate ratio can be computed for a sample of $N$ white dwarfs as:
\begin{align}
    \mathrm{PDF}^\star &= \frac{1}{N} \mathlarger{\mathlarger{\sum}}_{i=0}^{N} \frac{1}{\sqrt{2\pi}\sigma_\mathrm{{log(\dot{M}_\mathrm{Fe}/\dot{M}_\mathrm{Ca})}}} \label{eq:PDF_star}\\
    & \qquad\qquad \mathrm{exp}\left[{-\frac{1}{2}\left(\frac{\mathrm{log(\dot{M}_\mathrm{Fe}/\dot{M}_\mathrm{Ca})}-\mathrm{log(\dot{M}_\mathrm{Fe}/\dot{M}_\mathrm{Ca})_i}}{\sigma_\mathrm{log(\dot{M}_\mathrm{Fe}/\dot{M}_\mathrm{Ca})_i}}\right)^2}\right]. \nonumber
\end{align}
The error-corrected cumulative probability function ($\mathrm{CDF}^\star$) similarly follows from its integral as: 
\begin{equation}\label{eq:cdf}
    \mathrm{CDF}^\star = \frac{1}{2N} \mathlarger{\mathlarger{\sum}}_{i=0}^{N} \left[1+\mathrm{erf}\left(\frac{\mathrm{log(\dot{M}_\mathrm{Fe}/\dot{M}_\mathrm{Ca})}-\mathrm{log(\dot{M}_\mathrm{Fe}/\dot{M}_\mathrm{Ca})_i}}{\sqrt{2}\sigma_\mathrm{log(\dot{M}_\mathrm{Fe}/\dot{M}_\mathrm{Ca})_i}}\right)\right].
\end{equation}
We estimate asymmetric errors on $\mathrm{CDF}^\star$ by repeatedly sampling $N$ times from Eq. \ref{eq:PDF_star}. Larger errors on the accretion ratio make a cumulative distribution of $\dot{M}_\mathrm{Fe}/\dot{M}_\mathrm{Ca}$ appear more shallow than its true distribution. Because the reported errors on the photospheric abundances of white dwarfs far exceed those of FGK stars, we apply the same white dwarf error to the FGK sample when the two curves are compared. For this comparison, we calculate the Kolmogorov–Smirnov test statistic ($D^\star$) of their two error-corrected CDFs:
\begin{figure}
\centering
\includegraphics[width=\hsize]{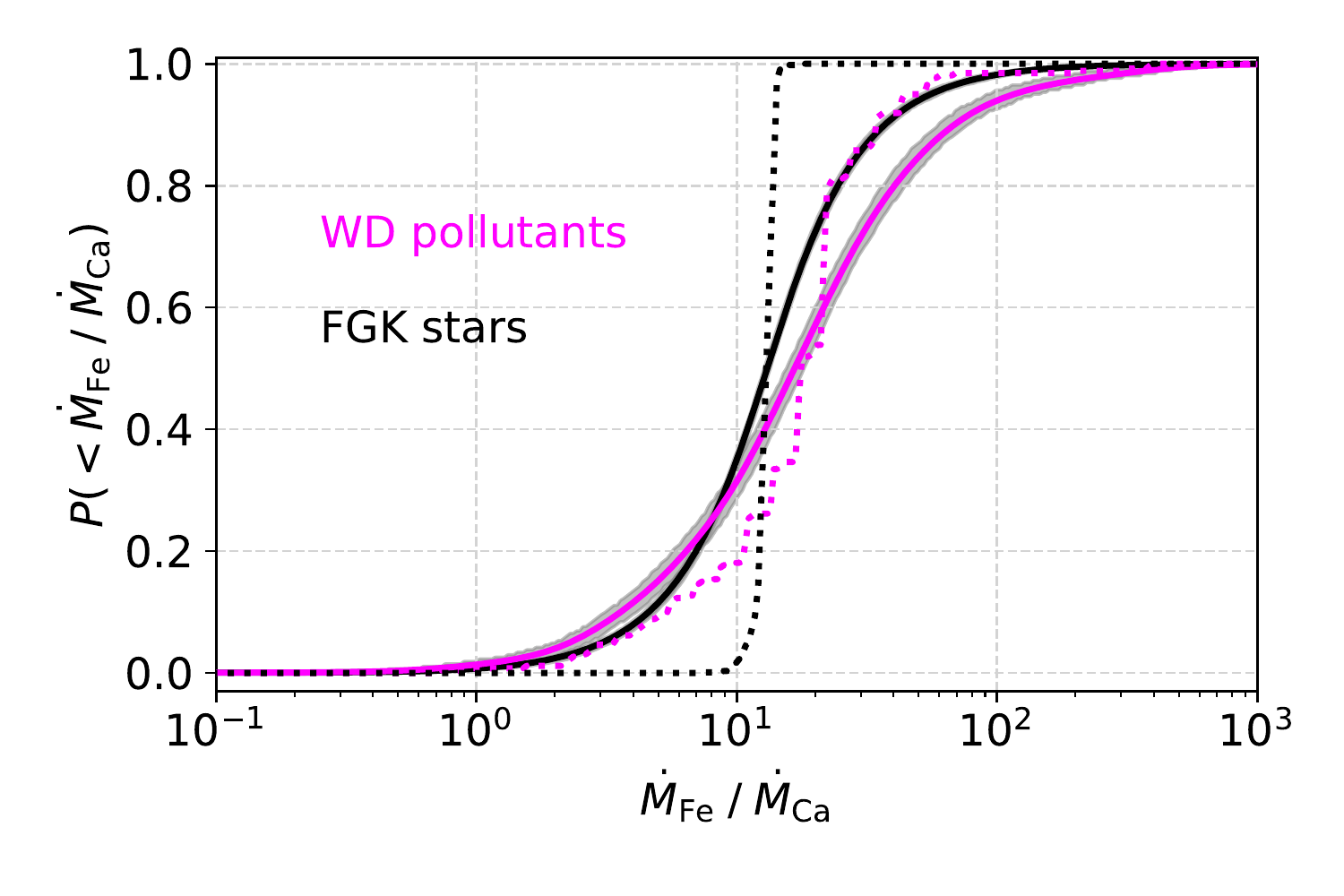}
\caption{Cumulative distribution of steady-state accretion rates (Eq. \ref{eq:Mdot_ss}) of iron relative to calcium, showing that white dwarf polluting material is iron-rich. The magenta curve corresponds to bulk white dwarf pollutant compositions (white dwarfs with $\tau_\mathrm{Fe}>\textrm{Myr}$), while the black curve corresponds to abundances of nearby FGK stars \citep{Brewer2016}. The dotted lines indicate reported values, and the solid lines are built from a sampling with white dwarf measurement errors ($\mathrm{CDF^*}$, see text, grey band indicates $1\sigma$ of this distribution). \label{fig:CaFe_vs_stars}}
\end{figure}
\begin{equation}\label{eq:D_star}
    D^\star = \mathrm{max}\left(|\mathrm{CDF}_\mathrm{FGK \; stars}^\star-\mathrm{CDF}_\mathrm{WD\; pollutants}^\star|\right).
\end{equation}
Next, the value of $D^\star$ is compared to the expected distribution of $D$ in the case that both subsamples (abundance ratios of FGK stars and white dwarfs) belong to the same population. To derive this distribution, we combine samples 1 and 2, and then resample $N_1$ and $N_2$ values of $\dot{M}_\mathrm{Fe}/\dot{M}_\mathrm{Ca}$ from this combined distribution, which yields a new value of $D$. This process is then repeated many times, and the probability that the two subsamples belong to the same broader population is identified as the fraction of samplings where $D>D^\star$.

The comparison between white dwarf pollutants and pristine material is shown in Fig. \ref{fig:CaFe_vs_stars}. First, we find that the Fe/Ca ratios of WD pollutants and FGK stars follow different distributions, with a significance exceeding $4\sigma$. Even corrected for errors, the white dwarf pollutants follow a significantly broader distribution of $\dot{M}_\mathrm{Fe}/\dot{M}_\mathrm{Ca}$, compared to the FGK stars. Because the plotted sample is limited to older white dwarfs ($\tau_\mathrm{Fe}>\textrm{Myr}$), that average over the accretion process, this indicates that the bulk content of white dwarf pollutants contains more compositional variation than is seen in the stellar sample, likely as a result of the collisional evolution of differentiated bodies prior to the accretion process \citep{Bonsor2020}. Alternatively, this widening could be caused by a spread in the accretion states of the white dwarfs (i.e. declining, build-up). Secondly, and more importantly, we find that the distribution of $\dot{M}_\mathrm{Fe}/\dot{M}_\mathrm{Ca}$ corresponding to white dwarf pollutants is off-set in the direction of increased core content. This direction cannot be explained by deviations from steady-state accretion, as a build-up state is unlikely for cool white dwarfs with helium-dominated atmospheres, and $\tau_\mathrm{Ca}>\tau_\mathrm{Fe}$. However, the preferential ejection of mantle fragments can increase the bound core mass fraction by up to 20\% (see Fig. \ref{fig:core_increase_BulkEarth}), which is enough to explain the off-set.

\subsection{Investigating the core-mantle accretion asynchronicity}\label{sect:evidence_asynchronous}
\begin{figure*}
\centering
\includegraphics[width=\hsize]{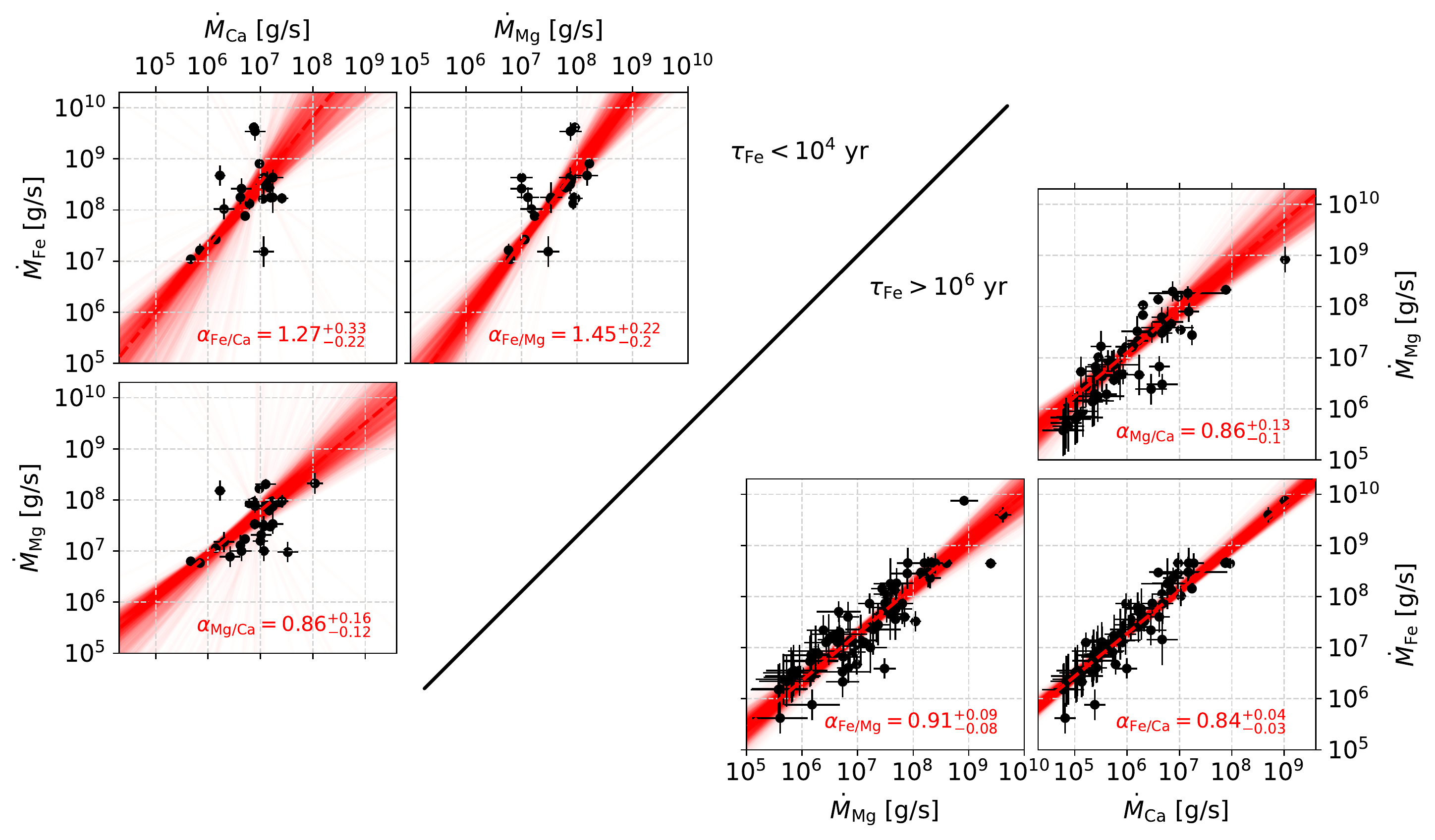}
\caption{Accretion rates of iron relative to lithophile elements (Mg, Ca) in our white dwarf sample. The rows separate stars with short (left, $<10^4$ yr) and long (right, $\textrm{>Myr}$) diffusion timescales. The red line indicates the best-fit log-log trend with slope $\alpha$ (Eq. \ref{eq:logMdot_trend}), and the transparent lines show the bootstrapped distribution. White dwarfs with short diffusion timescales have best-fit values between Fe and Mg or Ca of $\alpha \gtrsim 1$, while stars with long diffusion timescales have $\alpha \sim 1$. The fits are best constrained for older stars with long diffusion timescales due to the larger available sample.}\label{fig:mdot_scatter_fit}
\end{figure*}

Next, we investigate the second prediction, that core and mantle fragments accrete in a proportion that varies over time during a single accretion event. For this purpose, we schematically decompose the accretion rate of an element El at time $t$ into two physically motivated terms:
\begin{equation}\label{eq:Mdot_sketch}
    \dot{M}_\mathrm{El} \propto S\left(R_\mathrm{ast}\right) \times
    \mathrm{exp[-t/t_\mathrm{acc,El}]}.
\end{equation}
The first function $S(R_\mathrm{ast})$ refers to the influence of a pollutant's size on the accretion rate of an element, roughly proportional to $R_\mathrm{ast}^3$. This term affects all elements evenly, as a bigger asteroid contains more of every element. The second term indicates the time-dependent factor, where different elements can accrete over different timescales $\tau_\mathrm{acc,El}$. In our analysis, we set out to investigate the relative importance of these two terms. To do so, we fit the relationship between $\dot{M}_\mathrm{Fe}$ and $\dot{M}_\mathrm{Ca}$, and look for a relation of the form:
\begin{equation}\label{eq:logMdot_trend}
    \mathrm{log(\dot{M}_\mathrm{Fe})} = \alpha_\mathrm{Fe/Ca} \;  \mathrm{log(\dot{M}_\mathrm{Ca})} + \beta_\mathrm{Fe/Ca}.
\end{equation}
If the size-dependent factor $S$ dominates the accretion rate in Eq. \ref{eq:Mdot_sketch}, the scaling factor $\alpha_\mathrm{Fe/Ca}$ is expected to be close to unity. However, if all asteroids are the same size, and the fragments from the core accrete more rapidly than those from the mantle, the value of $\alpha$ is expected to be significantly greater for stars with diffusion timescales shorter than the duration of accretion, when compared to stars with long diffusion timescales, whose abundances average over the time-dependent factor $g$ in Eq. \ref{eq:Mdot_sketch}.

We use the module \texttt{scipy.odr} to calculate the best-fit line of Eq. \ref{eq:logMdot_trend} via orthogonal regression with weighed error bars. The errors on the slope and intercept are calculated via repeated fitting of white dwarf sets that are randomly sampled with replacement in a procedure called bootstrapping \citep[e.g.,][]{Montgomery2021}. The best-fit value of $\alpha$ is determined as the median of the bootstrapped distribution, and its errors are identified from the $15.8^\mathrm{th}$ and $84.1^\mathrm{st}$ percentiles. We split the sample of white dwarfs into two bins, depending on the diffusion timescale of iron. White dwarfs with short diffusion timescales measure ongoing accretion, while long diffusion times average over past accretion.

The results of this analysis are shown in Fig. \ref{fig:mdot_scatter_fit}. In the sample of white dwarfs with long diffusion timescales, the fits are well-constrained due to the large sample size, and are around $\alpha \simeq 1$, which is expected when the averaged rate of accretion during an event is determined by the size of pollutants. Interestingly, the best-fit values of $\alpha_\mathrm{Fe/Mg}$ and $\alpha_\mathrm{Fe/Ca}$ are higher in the sample with short diffusion timescales, while the best-fit value of $\alpha_\mathrm{Mg/Ca}$ does not show a similar difference. This observation is consistent with a scenario of asynchronous accretion, where core fragments accrete more rapidly onto the star than mantle fragments. However, the fits are more poorly constrained in the sample of white dwarfs with short diffusion timescales. With an extended sample of young white dwarfs with hydrogen-dominated atmospheres, a similar procedure could be used in the future to corroborate these findings.

\section{Discussion}\label{sect:discussion}

\subsection{Interpretation of photospheric abundances}
The suggested asynchronous accretion of core and mantle material means that the age and type of white dwarfs should be accounted for when interpreting their photospheric abundances. For young white dwarfs with hydrogen-dominated atmospheres and short diffusion timescales, their photospheric composition may vary during the accretion of a single body. The positive implication of this hypothesis is that observers and geologists can more easily study the exoplanetary geology of their pollutants, as every differentiated body goes through a core and mantle-rich accretion phase, and catastrophic collisions between asteroids are not required in order to observationally sample these layers. The flip side of asynchronous accretion is that a pollutant's bulk composition cannot be confidently inferred when the atmospheric diffusion timescale of a white dwarf is shorter than the timescale on which the accreted composition changes. In short, a young white dwarf with hydrogen-dominated envelope might sample the composition of a given layer of the pollutant, while older, helium-dominated white dwarfs with long diffusion timescales are more suited to infer the bulk composition of their pollutants. However, even old white dwarfs are not perfect spectrometers of white dwarf pollutants, as the preferential ejection of mantle fragments in a tidal disruption (Section \ref{subsect:ejection}) implies that the total material accreted by a white dwarf will be enriched in core material by up to 20\% when it accretes a differentiated body. In addition, older white dwarfs come with their own inherent difficulties of a more poorly constrained accretion state and history, as they could potentially have swallowed multiple objects in a single diffusion time \citep{Wyatt2014, Turner2020, Trierweiler2022}.

\subsection{Observational tests for asynchronous core-mantle accretion}
Accretion asynchronicities can be directly studied observationally with a sufficiently varied white dwarf sample with well-constrained and diverse abundances. While a sufficiently good sample does not yet exist, we suggest a procedure in Section \ref{sect:evidence_asynchronous} that can be used to study this process in the future. If the abundance ratios of core and mantle material vary over a characteristic time period during accretion, regression between the accretion rates of siderophile and lithophile elements will show a steep trend, but only for stars with diffusion times shorter than the typical accretion event. Using the current white dwarf sample, we find tentative evidence that the size range of pollutants is broad, and core material accretes more rapidly than mantle material. We note, however, that a larger sample of young white dwarfs with hydrogen-dominated atmospheres ($\sim$ 100 stars) is required to corroborate these findings. In addition, the identification of other siderophile elements like chromium in more white dwarfs would aid the comparison, as it will allow for an independent analysis of the same physical trend. In the future, upcoming large-scale spectroscopic surveys
(4MOST/WEAVE/DESI/SDSS-V) will greatly increase the number of known, young white dwarfs with photospheric pollution. Studying the asynchronicity of the accretion process will put much-needed observational constraints on the white dwarf accretion process.

\subsection{Asynchronous accretion in alternative models}\label{sect:alternative_models}
In this work, we argue that core and mantle fragments of differentiated asteroids are expected to accrete onto white dwarfs in proportions that vary over time. We note, however, that the details and magnitude of this asynchronicity are intimately linked with the accretion process itself, which is still imperfectly understood. Most current theoretical studies share the idea that accretion begins with the tidal disruption of a pollutant \citep{Malamud2020a, Malamud2020b, Malamud2021, Li2021, Trevascus2021, Hogg2021, Zhang2021, Veras2021a, Brouwers2022a}. We use energy arguments to show that such a disruption spreads core and mantle fragments to different orbits (see Fig. \ref{fig:disruption_geometries}), leading to the preferential ejection of mantle material. This asymmetric distribution forms the starting point of asynchronous accretion in our models. For the subsequent accretion process, we considered two simple scenarios, each of which has its limitations.

In our first model of collisional grind-down, the main simplification is that it only tracks catastrophic collisions, rather than following the full collisional evolution of child orbits. As discussed by \citet{Brouwers2022a}, it is possible that a more detailed tracking of the collision tree would yield either a reduced or increased asynchronicity between core and mantle fragments, or that this trend is altered when other sources of collisions are accounted for (e.g., gravitational stirring \citep{Li2021}, the Yarkovski effect \citep{Veras2015a, Veras2015b, Veras2020a}, Poynting-Robertson drag \citep{Rafikov2011a}). Finally, the fragment size distribution could differ between core and mantle fragments, as iron is denser, stronger and more ductile than typical mantle minerals. Our analysis of fragment scattering by a planet is yet more simple and just serves to illustrate the idea that core fragments are more susceptible to close encounters with the planet, rather than produce an exact time evolution.

It is also conceivable that accretion begins with a tidal disruption, but then proceeds via different channels than the ones studied in this work. For instance, in paper II, we describe how the rapid sublimation of ices can cause a distinct accretion asynchronicity where volatiles reach the star faster than refractory components. Within the context of core-mantle accretion, \citet{Hogg2021, Zhang2021} suggest diamagnetic and Alfvén-wave drag as mechanisms to circularize and accrete magnetized fragments onto the star. In these cases too, however, there are asymmetries between more and less strongly magnetized fragments, which would translate to a conceptually similar, though quantitatively different core-mantle asynchronicity. A third possibility is that fragments circularize by interactions with a pre-existing disc, as suggested by \citet{Malamud2021}. In this case, the accretion times of fragments are largely determined by their semi-major axes, which also vary between core and mantle fragments (see Fig. \ref{fig:disruption_geometries}). As such, we argue that while the asynchronicity of core-mantle accretion is for now difficult to constrain theoretically, it is likely to affect relative white dwarf abundances in a wide range of accretion scenarios that involve a tidal disruption on a highly-eccentric orbit.

Asynchronous core-mantle accretion may also occur in the very different scenario where large objects circularize before they disrupt. Such a scenario might play out around some white dwarfs, considering the transits of disintegrating planetesimals seen in some systems \citep{Vanderburg2015, Manser2019b, Farihi2022, Budaj2022}. In the example of WD 1145+017, these transits are best modelled by differentiated planetesimals, whose mantles are shedding material, while their denser cores remain intact initially \citep{Veras2017, Duvvuri2020}.

\section{Summary and conclusions}\label{sect:conclusions}
Polluted white dwarfs with multiple identified photospheric elements are often used to infer the composition of the planetary bodies that accrete onto them \citep[e.g.,][]{Zuckerman2007, Klein2010, Farihi2011, Xu2014b, Xu2017, Xu2019, Hollands2018, Harrison2018, Harrison2021b, Swan2019b, Putirka2021, Buchan2022}. In this work and in an accompanying paper, we argue based on theory and observations that the composition of the material accreting onto a white dwarf may vary with time during the accretion of a single planetary body. Consequently, photospheric abundance ratios of white dwarfs can fluctuate during a single accretion event, and the abundances of white dwarfs do not necessarily reflect the bulk composition of their pollutants, especially for young stars with hydrogen-dominated atmospheres. The potential consequences are particularly important for differentiated bodies, whose cores and mantles disrupt into distinct groups of fragments that follow different orbits, and could accrete over varying periods of time. The models presented here make the following theoretical predictions:
\begin{enumerate}
    \item If a white dwarf accretes a core-mantle differentiated pollutant, the material accreted by white dwarfs will be enriched in core material by up to 20\% due to the ejection of a larger portion of the mantle during the tidal disruption.

    \item Both a collisional model and fragment scattering by a planet predict that accretion begins with an iron-rich phase, followed by a more Ca and Mg-rich second phase. This variation is caused by the geometry of a differentiated body during a tidal disruption, which implies that core fragments cluster around the centre of the disc that forms, while mantle fragments occupy both inner and outer orbits, and take longer to accrete on average.
\end{enumerate}
Analysis of the known polluted white dwarfs provides tentative observational support in favour of both of these predictions: 
\begin{enumerate}
    \item There are more white dwarfs accreting material with high Fe/Ca ratios than low Fe/Ca, assuming that relative sinking timescales for Ca and Fe are accurate. This can be interpreted as evidence for the ejection of mantle material when a differentiated pollutant tidally disrupts.

    \item We find tentative evidence that the accretion rate of iron declines more quickly than that of magnesium and calcium, as shown by the steep trend between $\dot{M}_\mathrm{Fe},\dot{M}_\mathrm{Mg}$ and $\dot{M}_\mathrm{Fe},\dot{M}_\mathrm{Ca}$ in the subsample of white dwarfs with short ($<10^4$ yr) diffusion timescales. This observation is consistent with the scenario that white dwarf accretion begins with a rapid, iron-rich accretion phase, followed by slower, more Ca and Mg-rich accretion.
\end{enumerate}
%

\section*{Data availability}
The simulation data that support the findings of this study are available upon request from the corresponding author, Marc G. Brouwers.

\section*{Acknowledgements}
We would like to thank Andrew Buchan, Laura Rogers, Elliot Lynch and Simon Blouin for useful discussions that helped shape this paper. Marc G. Brouwers acknowledges the support of a Royal Society Studentship, RG 16050. Amy Bonsor acknowledges support from a Royal Society Dorothy Hodgkin Research Fellowship (grant number DH150130) and a Royal Society University Research Fellowship (grant number URFR1211421).

\bibliographystyle{mnras}
\bibliography{circularisation}

\clearpage
\begin{appendix}

\section{Neglection of asteroid rotation on tidal disc geometry}\label{appendix:sg_rot}
The predicted asynchronicity between the accretion of core and mantle fragments due to collisions or scattering is caused by their spreading to distinct orbital zones after a tidal disruption. In our calculation of the fragment orbits (Eq. \ref{eq:a_new_2}), we neglected the effect of self-gravity and the rotation of the disrupted object. In this appendix, we validate that these contributions are indeed negligible in the context of white dwarf pollution by asteroids or minor planets. To account for the energy of rotation ($\epsilon_\mathrm{rot,i}$) and self-gravity ($u_\mathrm{SG,i}$), we modify the total fragment energy ($\epsilon_\mathrm{i}$) in Eq. 5 of \citet{Brouwers2022a}:
\begin{subequations}
\begin{align}
    \epsilon_\mathrm{i} &=  u_\mathrm{\star G,i} + u_\mathrm{SG,i} + \epsilon_\mathrm{k,i} + \epsilon_\mathrm{rot,i} \\
    &= - \frac{GM_\mathrm{WD}}{r_\mathrm{i}} - \frac{GM\left(<R_\mathrm{i}\right)}{R_\mathrm{i}}+\frac{1}{2}\left| GM_\mathrm{WD}\left(\frac{2}{r_\mathrm{B}}-\frac{1}{a_0}\right)\mathbf{\hat{\theta}} + \mathbf{\omega_\mathrm{rot}^2} \times \mathbf{R_\mathrm{i}}^2\right|, \label{eq:e_width}
\end{align}
\end{subequations}
where $\mathbf{\omega_\mathrm{rot}}$ is the rotation vector of the disrupted object and $\mathbf{R_\mathrm{i}}$ is the vector distance of a fragment to the object's centre. To investigate at what size the two new terms begin to dominate differences between fragment orbits, we evaluate the transition size $R_\mathrm{transition}$ where their radial derivatives along a line through the star and the disrupting object's centre are equal:
\begin{equation}
    \frac{d}{dR_\mathrm{i}} \left(u_\mathrm{\star G,i} + \epsilon_\mathrm{k,i} \right) = \frac{d}{dR_\mathrm{i}} \left(u_\mathrm{SG,i} + \epsilon_\mathrm{rot,i} \right),
\end{equation}
where $\mathbf{R_\mathrm{i}} = (r_\mathrm{i} - r_\mathrm{b}) \mathbf{\hat{r}}$ because we consider a radial line. The transition size is smallest if the object's rotation is equal to the critical rate $\omega_\mathrm{crit} = \sqrt{4\pi G\rho/3}$ \citep{Pravec2000}, in which case it evaluates to:
\begin{subequations}
\begin{align}
    R_\mathrm{transition} &= \frac{M_\mathrm{WD}}{4\pi G \rho r_\mathrm{B}^2} \qquad \mathrm{if \;\; \mathbf{\omega_\mathrm{rot}} =  \omega_\mathrm{crit} \mathbf{\hat{z}}}\\
    &=  3.9\cdot10^4\;\mathrm{km}\; \left(\frac{\rho}{5 \; \mathrm{g/cm^3}}\right)^{-1} \left(\frac{r_\mathrm{B}}{\mathrm{R_\odot}}\right)^{-1} \left(\frac{M_\mathrm{WD}}{0.6\;\mathrm{M_\odot}}\right),
\end{align}
\end{subequations}
which corresponds to a size in between Uranus and Jupiter, far larger than the asteroids and minor planets considered in this work. Therefore, we conclude that rotation and self-gravity do not affect the general shape of the tidal disc. This is consistent with the SPH simulations by \citet{Malamud2020a, Malamud2020b}, where Earth-mass planets were shown to tidally disrupt into tidal discs with distinct core and mantle fragment zones. 

\section{Filling timescale of a tidal disc}\label{appendix:spreading}
\begin{figure}
\centering
\includegraphics[width=\hsize]{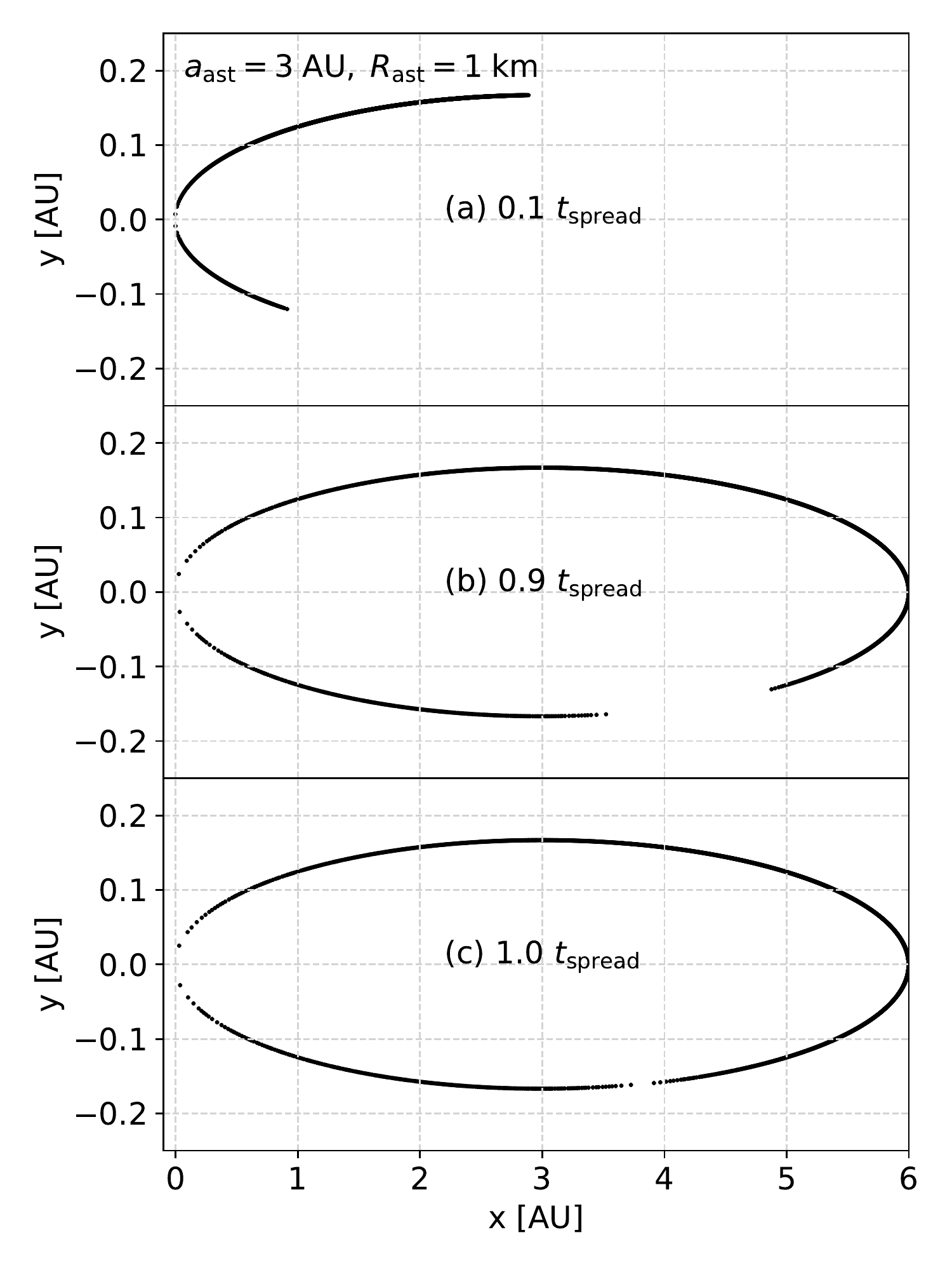}
\caption{Snapshots indicating the spreading of fragments in a tidal disc after the disruption of a small asteroid ($a_\mathrm{ast}=3\; \mathrm{AU}, \; R_\mathrm{ast}=1\; \mathrm{km}$). The inner and outer fragments re-align after nearly precisely one spreading timescale, as defined by Eq. \ref{eq:t_spread}. \label{fig:spreading_timescale}}
\end{figure}
When an asteroid tidally disrupts, its fragments are initially clustered around the same true anomaly, and some time is required to spread them out into a fully formed tidal disc. The validity of the accretion models discussed in Section \ref{sect:accretion} depends on the requirement that the fragments completely spread out to fill a disc before other accretion processes (scattering, collisions, sublimative erosion) become important. In this appendix, we provide a short derivation of this spreading timescale. Our derivation is most comparable to calculations by \citet{Nixon2020} and follows similar arguments to those presented by \citet{Veras2014}, although our derivation is applicable to both dispersive and non-dispersive disruptions.

We begin by specifying the semi-major axis of ($a_\mathrm{inner}$) and outer ($a_\mathrm{outer}$) fragments from Eq. \ref{eq:a_new_2}. The fragments break off in a range of distances from the central star between $r_\mathrm{B} - R_\mathrm{ast}$ and $r_\mathrm{B} + R_\mathrm{ast}$, which yields:
\begin{subequations}
\begin{align}
    a_\mathrm{inner} &= a_\mathrm{ast} \left(1+\frac{2a_\mathrm{ast}R_\mathrm{ast}}{r_\mathrm{B}\left(r_\mathrm{B}-R_\mathrm{ast}\right)}\right)^{-1}, \\
    a_\mathrm{outer} &= a_\mathrm{ast} \left(1-\frac{2a_\mathrm{ast}R_\mathrm{ast}}{r_\mathrm{B}\left(r_\mathrm{B}+R_\mathrm{ast}\right)}\right)^{-1}.
\end{align}
\end{subequations}
The total orbital width $(\Delta a)$ of the tidal disc that forms is thus:
\begin{subequations}
\begin{align}
    \Delta a &= a_\mathrm{outer} - a_\mathrm{inner} \\
    &\simeq \frac{4 a_\mathrm{ast}^2R_\mathrm{ast}}{r_\mathrm{B}^2},
\end{align}
\end{subequations}
where we used a Maclaurin expansion around $R_\mathrm{ast}/r_\mathrm{B} \ll 1$, which is always valid for disruptions around white dwarfs. Next, the derivative of the orbital time $P = 2\pi \sqrt{a/G M_\mathrm{WD}}$ can be used to specify the corresponding spread in orbital periods $\Delta P$:
\begin{subequations}
\begin{align}
    \Delta P &= \frac{dP}{da} \Delta a \\
    &= 3\pi \sqrt{\frac{a_\mathrm{ast}}{G M_\mathrm{WD}}} \Delta a.
\end{align}
\end{subequations}
The spreading timescale $t_\mathrm{spread}$ can be defined as the time required for and outer particle to return to the same true anomaly. The number of orbits required is $N_\mathrm{spread}=P/\Delta P$, so the spreading timescale can be written as:
\begin{subequations}
\begin{align}
    t_\mathrm{spread} &= \frac{P^2}{\Delta P} \\
    &= \frac{\pi r_\mathrm{B}^2}{3R_\mathrm{ast}}\sqrt{\frac{a_\mathrm{ast}}{G M_\mathrm{WD}}} \label{eq:t_spread} \\
    &= 120 \;\mathrm{yr}\; \left(\frac{a_\mathrm{ast}}{3\;\mathrm{AU}}\right)^\frac{1}{2} \left(\frac{R_\mathrm{ast}}{\mathrm{km}}\right)^{-1}\left(\frac{r_\mathrm{B}}{\mathrm{R_\odot}}\right)^2 \left(\frac{M_\mathrm{WD}}{0.6\;\mathrm{M_\odot}}\right)^{-\frac{1}{2}},
\end{align}
\end{subequations}
which is much shorter than typical accretion timescales, but might be comparable to the initial accretion phase. In Fig. \ref{fig:spreading_timescale}, we validate Eq. \ref{eq:t_spread} with a visual comparison to the position of fragments in a tidal disc at three snapshots in time. The orbital elements of the fragments are given by Eq. \ref{eq:a_new_2} and their true anomalies are exactly specified as a function of time by Kepler's laws. The filling of the tidal disc agrees well with Eq. \ref{eq:t_spread} and and outer fragments re-align after nearly exactly one spreading timescale.

\section{White dwarf sample used in this work}

\onecolumn
\begin{center}\label{table:sample}
\small
\begin{longtable}{l c c c c c c c c}
\caption{Properties and abundances of white dwarfs in the sample, ordered by type and temperature} \\\hline
\hline
\hline
System & Type & $T_\mathrm{eff}$ [K] & Mass [$\mathrm{M_\odot}$] & log($g / \mathrm{ms}^{-2}$) & log(Fe/Hx) & log(Mg/Hx) & log(Si/Hx) & log(Ca/Hx)\\ 
\hline
\endhead
\hline
{{\bfseries \tablename\ \thetable{} -- continued on next page}}
\endfoot
\hline
\endlastfoot
GaiaJ2100+2122${}^{(hh,\overline{dd})}$ & H & 25565.0 & 0.693 & 8.1 & -4.96$\pm$0.1 & -5.08$\pm$0.1 & -5.13$\pm$0.12 & -6.23$\pm$0.13\\
GaiaJ2100+2122${}^{(hh,\overline{dd})}$ & H & 25565.0 & 0.693 & 8.1 & -4.96$\pm$0.2 & -5.12$\pm$0.2 & -5.12$\pm$0.2 & -6.23$\pm$0.2\\
PG0843+516${}^{(f,\overline{u})}$ & H & 22412.0 & 0.577 & 7.902 & -3.84$\pm$0.18 & -4.82$\pm$0.2 & -4.59$\pm$0.12 & -6.26$\pm$0.2\\
GALEX1931+0117${}^{(q,\overline{d})}$ & H & 21457.0 & 0.573 & 7.9 & -4.43$\pm$0.09 & -4.42$\pm$0.06 & -4.24$\pm$0.07 & -6.11$\pm$0.05\\
SDSSJ1228+1040${}^{(j,\overline{e})}$ & H & 20900.0 & 0.73 & 8.15 & -5.2$\pm$0.3 & -5.2$\pm$0.2 & -5.2$\pm$0.2 & -5.94$\pm$0.2\\
PG1015+161${}^{(e,\overline{u})}$ & H & 19226.0 & 0.642 & 8.04 & -4.92$\pm$0.2 & -5.6$\pm$0.2 & -5.42$\pm$0.21 & -6.4$\pm$0.2\\
SDSSJ1043+0855${}^{(j,\overline{t})}$ & H & 18320.0 & 0.65 & 8.05 & -6.15$\pm$0.3 & -5.11$\pm$0.2 & -5.33$\pm$0.5 & -5.96$\pm$0.2\\
WD0611-6931${}^{(hh,\overline{dd})}$ & H & 17750.0 & 0.702 & 8.14 & -3.77$\pm$0.1 & -4.67$\pm$0.1 & -4.6$\pm$0.21 & -6.19$\pm$0.1\\
HE0106-3253${}^{(d,a,\overline{u})}$ & H & 17350.0 & 0.62 & 8.12 & -4.7$\pm$0.06 & -5.57$\pm$0.2 & -5.48$\pm$0.05 & -5.93$\pm$0.11\\
GD56${}^{(b,\overline{u})}$ & H & 15270.0 & 0.67 & 8.09 & -5.44$\pm$0.2 & -5.55$\pm$0.2 & -5.69$\pm$0.2 & -6.86$\pm$0.2\\
WD1145+288${}^{(d,l,\overline{u})}$ & H & 12140.0 & 0.685 & 8.14 & -5.43$\pm$0.2 & -6.0$\pm$0.2 & <-4.7 & -6.88$\pm$0.08\\
G29-38${}^{(q,\overline{m})}$ & H & 11800.0 & 0.85 & 8.4 & -5.9$\pm$0.1 & -5.77$\pm$0.13 & -5.6$\pm$0.17 & -6.58$\pm$0.12\\
WD2105-820${}^{(aa,\overline{v})}$ & H & 10890.0 & 0.86 & 8.41 & -6.0$\pm$0.2 & -6.0$\pm$0.2 & <-5.5 & -8.2$\pm$0.1\\
GaiaJ1814-7355${}^{(hh,\overline{dd})}$ & H & 10190.0 & 0.58 & 7.996 & -6.06$\pm$0.19 & -6.14$\pm$0.08 & - & -7.22$\pm$0.15\\
WDJ1814-7354${}^{(dd,\overline{z})}$ & H & 10090.0 & 0.59 & 8.0 & -6.06$\pm$0.19 & -6.14$\pm$0.08 & <-6.0 & -7.22$\pm$0.15\\
WD2115-560${}^{(aa,\overline{v})}$ & H & 9600.0 & 0.58 & 7.97 & -6.4$\pm$0.1 & -6.4$\pm$0.1 & -6.2$\pm$0.1 & -7.4$\pm$0.1\\
WD1257+278${}^{(u, b,\overline{i})}$ & H & 8609.0 & 0.73 & 8.24 & -7.47$\pm$0.09 & -7.49$\pm$0.08 & - & -8.38$\pm$0.06\\
WD0354+463${}^{(u, b,\overline{i})}$ & H & 8240.0 & 0.57 & 7.96 & -7.13$\pm$0.11 & -6.7$\pm$0.05 & - & -8.2$\pm$0.03\\
NLTT25792${}^{(t,u,\overline{i})}$ & H & 7903.0 & 0.618 & 8.04 & -7.16$\pm$0.04 & -7.24$\pm$0.05 & - & -8.07$\pm$0.06\\
G166-58${}^{(b,\overline{u})}$ & H & 7390.0 & 0.58 & 7.99 & -8.22$\pm$0.13 & -8.06$\pm$0.05 & <-8.2 & -9.33$\pm$0.08\\
WD1455+298${}^{(u,\overline{i})}$ & H & 7383.0 & 0.589 & 7.97 & -8.4$\pm$0.08 & -8.03$\pm$0.06 & - & -9.51$\pm$0.03\\
G74-7${}^{(u,\overline{i})}$ & H & 7306.0 & 0.572 & 8.06 & -8.03$\pm$0.09 & -7.79$\pm$0.06 & - & -9.05$\pm$0.04\\
WD2157-574${}^{(aa,\overline{v})}$ & H & 7010.0 & 0.63 & 8.06 & -7.3$\pm$0.1 & -7.0$\pm$0.1 & -7.0$\pm$0.1 & -8.1$\pm$0.1\\
NLTT 6390${}^{(t,\overline{h})}$ & H & 6040.0 & 0.53 & 7.9 & -8.57$\pm$0.11 & -8.66$\pm$0.2 & - & -10.0$\pm$0.04\\
NLTT 1675${}^{(t,\overline{h})}$ & H & 6020.0 & 0.61 & 8.04 & -8.63$\pm$0.13 & -8.56$\pm$0.12 & - & -9.53$\pm$0.03\\
NLTT43806${}^{(e,\overline{c})}$ & H & 5838.0 & 0.704 & 8.186 & -7.8$\pm$0.17 & -7.1$\pm$0.13 & -7.2$\pm$0.14 & -7.9$\pm$0.19\\
NLTT 19686${}^{(y,\overline{p})}$ & H & 5230.0 & 0.54 & 7.93 & -8.93$\pm$0.14 & -8.2$\pm$0.3 & - & -8.7$\pm$0.04\\
WD1536+520${}^{(o,\overline{s})}$ & He & 20800.0 & 0.58 & 7.96 & -4.5$\pm$0.15 & -4.06$\pm$0.15 & -4.32$\pm$0.15 & -5.28$\pm$0.15\\
SDSSJ0845+2257${}^{(v,\overline{n})}$ & He & 19780.0 & 0.679 & 8.18 & -4.6$\pm$0.2 & -4.7$\pm$0.15 & -4.8$\pm$0.3 & -5.95$\pm$0.1\\
GaiaJ0644-0352${}^{(hh,\overline{dd})}$ & He & 18350.0 & 0.704 & 8.18 & -6.46$\pm$0.1 & -5.74$\pm$0.1 & -5.98$\pm$0.1 & -6.77$\pm$0.1\\
GaiaJ0644-0352${}^{(hh,\overline{dd})}$ & He & 18350.0 & 0.704 & 8.18 & -6.46$\pm$0.16 & -5.75$\pm$0.15 & -5.98$\pm$0.1 & -6.74$\pm$0.18\\
SDSSJ2047-1259${}^{(bb,cc,\overline{y})}$ & He & 17970.0 & 0.617 & 8.04 & -6.4$\pm$0.2 & -5.6$\pm$0.1 & -5.6$\pm$0.1 & -6.9$\pm$0.1\\
GD61${}^{(c,\overline{k})}$ & He & 17280.0 & 0.71 & 8.2 & -7.6$\pm$0.07 & -6.69$\pm$0.05 & -6.82$\pm$0.04 & -7.9$\pm$0.06\\
GD424${}^{(bb,\overline{x})}$ & He & 16560.0 & 0.77 & 8.25 & -5.53$\pm$0.12 & -5.15$\pm$0.04 & -5.29$\pm$0.04 & -6.15$\pm$0.05\\
GD378${}^{(ee,\overline{bb})}$ & He & 15620.0 & 0.551 & 7.93 & -7.51$\pm$0.36 & -7.44$\pm$0.2 & -7.49$\pm$0.12 & -8.7$\pm$0.76\\
G241-6${}^{(p,n,\overline{j})}$ & He & 15300.0 & 0.71 & 8.0 & -6.82$\pm$0.14 & -6.26$\pm$0.1 & -6.62$\pm$0.2 & -7.3$\pm$0.2\\
WD1551+175${}^{(d,o,n,\overline{u})}$ & He & 14756.0 & 0.57 & 8.02 & -6.6$\pm$0.1 & -6.29$\pm$0.05 & -6.33$\pm$0.1 & -6.93$\pm$0.07\\
WD2207+121${}^{(m,\overline{u})}$ & He & 14752.0 & 0.57 & 7.97 & -6.46$\pm$0.13 & -6.15$\pm$0.1 & -6.17$\pm$0.11 & -7.4$\pm$0.08\\
WD1145+017${}^{(k,\overline{w})}$ & He & 14500.0 & 0.656 & 8.11 & -5.61$\pm$0.2 & -5.91$\pm$0.2 & -5.89$\pm$0.2 & -7.0$\pm$0.2\\
WD1425+540${}^{(n,\overline{r})}$ & He & 14490.0 & 0.56 & 7.95 & -8.15$\pm$0.14 & -8.16$\pm$0.2 & -8.03$\pm$0.31 & -9.26$\pm$0.1\\
HS2253+8023${}^{(ee,\overline{b})}$ & He & 14400.0 & 0.84 & 8.4 & -6.17$\pm$0.17 & -6.1$\pm$0.14 & -6.27$\pm$0.13 & -6.99$\pm$0.11\\
SDSSJ0738+1835${}^{(i,\overline{g})}$ & He & 13950.0 & 0.841 & 8.4 & -4.98$\pm$0.09 & -4.68$\pm$0.07 & -4.9$\pm$0.16 & -6.23$\pm$0.15\\
GALEXJ2339${}^{(ee,\overline{bb})}$ & He & 13735.0 & 0.548 & 7.93 & -6.99$\pm$0.3 & -6.58$\pm$0.14 & -6.59$\pm$0.08 & -8.03$\pm$0.75\\
GD40${}^{(m,\overline{j})}$ & He & 13594.0 & 0.6 & 8.02 & -6.47$\pm$0.12 & -6.2$\pm$0.16 & -6.44$\pm$0.3 & -6.9$\pm$0.2\\
SDSSJ1242+5226${}^{(w,\overline{o})}$ & He & 13000.0 & 0.59 & 8.0 & -5.9$\pm$0.15 & -5.26$\pm$0.15 & -5.3$\pm$0.06 & -6.53$\pm$0.1\\
WD1232+563${}^{(m,\overline{u})}$ & He & 11787.0 & 0.77 & 8.3 & -6.45$\pm$0.11 & -6.09$\pm$0.05 & -6.36$\pm$0.13 & -7.69$\pm$0.05\\
WD1350-162${}^{(aa,\overline{v})}$ & He & 11640.0 & 0.6 & 8.02 & -7.1$\pm$0.1 & -6.8$\pm$0.1 & -7.3$\pm$0.2 & -8.7$\pm$0.1\\
PG1225-079${}^{(g,\overline{f,l})}$ & He & 10800.0 & 0.58 & 8.0 & -7.42$\pm$0.07 & -7.5$\pm$0.2 & -7.45$\pm$0.1 & -8.06$\pm$0.03\\
WD0446-255${}^{(aa,\overline{v})}$ & He & 10120.0 & 0.58 & 8.0 & -6.9$\pm$0.1 & -6.6$\pm$0.1 & -6.5$\pm$0.1 & -7.4$\pm$0.1\\
GD362${}^{(r,s,\overline{a})}$ & He & 10057.0 & 0.551 & 7.95 & -5.65$\pm$0.1 & -5.98$\pm$0.25 & -5.84$\pm$0.3 & -6.24$\pm$0.1\\
WD0449-259${}^{(aa,\overline{v})}$ & He & 9850.0 & 0.61 & 8.04 & -7.9$\pm$0.2 & -8.3$\pm$0.4 & <-7.3 & -9.1$\pm$0.1\\
WD2216-657${}^{(aa,\overline{v})}$ & He & 9120.0 & 0.61 & 8.05 & -8.0$\pm$0.2 & -7.1$\pm$0.1 & <-7.0 & -9.0$\pm$0.1\\
SDSSJ0956+5912${}^{(z,\overline{q})}$ & He & 8843.0 & 0.683 & 8.168 & -6.2$\pm$0.1 & -5.2$\pm$0.1 & - & -7.1$\pm$0.1\\
SDSSJ1340+2702${}^{(z,\overline{q})}$ & He & 8413.0 & 0.6 & 8.0 & -5.6$\pm$0.1 & -5.1$\pm$0.1 & - & -7.0$\pm$0.1\\
WD0122-227${}^{(aa,\overline{v})}$ & He & 8380.0 & 0.61 & 8.06 & -8.5$\pm$0.2 & -8.5$\pm$0.4 & <-7.6 & -10.1$\pm$0.1\\
SDSSJ0956+5912${}^{(ff,\overline{cc})}$ & He & 8100.0 & 0.59 & 8.02 & -6.9$\pm$0.1 & -5.5$\pm$0.1 & -5.7$\pm$0.2 & -7.3$\pm$0.05\\
SDSSJ0046+2717${}^{(z,\overline{q})}$ & He & 8053.0 & 0.879 & 8.465 & -6.3$\pm$0.3 & -5.4$\pm$0.2 & - & -7.2$\pm$0.2\\
SDSSJ1038-0036${}^{(z,\overline{q})}$ & He & 7996.0 & 0.6 & 8.0 & -6.9$\pm$0.1 & -6.3$\pm$0.1 & - & -7.4$\pm$0.1\\
SDSSJ1158+4712${}^{(z,\overline{q})}$ & He & 7840.0 & 0.6 & 8.0 & -6.9$\pm$0.2 & -6.7$\pm$0.1 & - & -8.3$\pm$0.1\\
SDSSJ0744+2701${}^{(z,\overline{q})}$ & He & 7829.0 & 0.676 & 8.16 & -7.0$\pm$0.2 & -6.7$\pm$0.2 & - & -7.8$\pm$0.2\\
SDSSJ1041+3432${}^{(z,\overline{q})}$ & He & 7728.0 & 0.769 & 8.301 & -6.8$\pm$0.2 & -6.8$\pm$0.2 & - & -8.0$\pm$0.2\\
SDSSJ1308+0957${}^{(z,\overline{q})}$ & He & 7692.0 & 0.6 & 8.0 & -6.9$\pm$0.3 & -7.0$\pm$0.2 & - & -8.1$\pm$0.2\\
SDSSJ1356+0236${}^{(z,\overline{q})}$ & He & 7662.0 & 0.595 & 8.028 & -6.7$\pm$0.2 & -6.3$\pm$0.1 & - & -8.0$\pm$0.1\\
SDSSJ0816+2330${}^{(z,\overline{q})}$ & He & 7642.0 & 0.6 & 8.0 & -6.5$\pm$0.3 & -6.0$\pm$0.3 & - & -7.6$\pm$0.3\\
SDSSJ1234+5208${}^{(z,\overline{q})}$ & He & 7627.0 & 0.637 & 8.098 & -6.3$\pm$0.1 & -5.7$\pm$0.1 & - & -7.5$\pm$0.1\\
SDSSJ1038-0036${}^{(ff,\overline{cc})}$ & He & 7560.0 & 0.61 & 8.06 & -7.4$\pm$0.1 & -6.6$\pm$0.1 & -6.4$\pm$0.2 & -7.6$\pm$0.05\\
SDSSJ0946+2024${}^{(z,\overline{q})}$ & He & 7540.0 & 0.6 & 8.0 & -6.9$\pm$0.2 & -6.5$\pm$0.2 & - & -8.0$\pm$0.2\\
SDSSJ2319+3018${}^{(z,\overline{q})}$ & He & 7478.0 & 0.6 & 8.0 & -7.3$\pm$0.3 & -7.2$\pm$0.3 & - & -8.5$\pm$0.3\\
SDSSJ0252+0054${}^{(z,\overline{q})}$ & He & 7478.0 & 0.596 & 8.031 & -7.2$\pm$0.2 & -7.1$\pm$0.2 & - & -8.5$\pm$0.2\\
SDSSJ1319+3641${}^{(z,\overline{q})}$ & He & 7464.0 & 0.763 & 8.294 & -7.4$\pm$0.3 & -7.2$\pm$0.2 & - & -8.6$\pm$0.3\\
SDSSJ1150+4928${}^{(z,\overline{q})}$ & He & 7417.0 & 0.694 & 8.189 & -7.9$\pm$0.5 & -7.5$\pm$0.5 & - & -9.1$\pm$0.3\\
SDSSJ1320+0204${}^{(z,\overline{q})}$ & He & 7356.0 & 0.6 & 8.0 & -7.1$\pm$0.3 & -7.0$\pm$0.3 & - & -8.3$\pm$0.3\\
SDSSJ1507+4034${}^{(z,\overline{q})}$ & He & 7304.0 & 0.6 & 8.0 & -6.8$\pm$0.2 & -6.3$\pm$0.1 & - & -8.0$\pm$0.2\\
SDSSJ1144+3720${}^{(z,\overline{q})}$ & He & 7280.0 & 0.446 & 7.768 & -7.6$\pm$0.4 & -7.0$\pm$0.2 & - & -8.5$\pm$0.2\\
SDSSJ0901+0752${}^{(z,\overline{q})}$ & He & 7263.0 & 0.57 & 7.989 & -6.1$\pm$0.2 & -5.8$\pm$0.2 & - & -7.1$\pm$0.2\\
SDSSJ0902+1004${}^{(z,\overline{q})}$ & He & 7250.0 & 0.6 & 8.0 & -8.19$\pm$0.22 & -7.29$\pm$0.24 & - & -8.25$\pm$0.22\\
SDSSJ0143+0113${}^{(z,\overline{q})}$ & He & 7229.0 & 0.687 & 8.178 & -7.0$\pm$0.1 & -6.6$\pm$0.1 & - & -8.2$\pm$0.1\\
SDSSJ2352+3344${}^{(z,\overline{q})}$ & He & 7200.0 & 0.6 & 8.0 & -7.1$\pm$0.3 & -6.9$\pm$0.2 & - & -8.3$\pm$0.2\\
SDSSJ1612+3534${}^{(z,\overline{q})}$ & He & 7181.0 & 0.6 & 8.0 & -7.3$\pm$0.5 & -7.3$\pm$0.3 & - & -8.5$\pm$0.5\\
SDSSJ1058+3143${}^{(z,\overline{q})}$ & He & 7173.0 & 0.675 & 8.159 & -7.7$\pm$0.2 & -7.4$\pm$0.2 & - & -9.0$\pm$0.1\\
SDSSJ1149+0519${}^{(z,\overline{q})}$ & He & 7173.0 & 0.525 & 7.914 & -7.5$\pm$0.2 & -7.3$\pm$0.1 & - & -8.2$\pm$0.1\\
SDSSJ0842+1406${}^{(z,\overline{q})}$ & He & 7075.0 & 0.567 & 7.985 & -7.4$\pm$0.1 & -7.2$\pm$0.1 & - & -8.4$\pm$0.1\\
SDSSJ1443+5833${}^{(z,\overline{q})}$ & He & 7061.0 & 0.6 & 8.0 & -7.3$\pm$0.2 & -7.1$\pm$0.1 & - & -8.5$\pm$0.2\\
SDSSJ1405+1549${}^{(z,\overline{q})}$ & He & 7055.0 & 0.621 & 8.073 & -7.3$\pm$0.1 & -6.9$\pm$0.1 & - & -8.5$\pm$0.1\\
SDSSJ1445+0913${}^{(z,\overline{q})}$ & He & 7035.0 & 0.6 & 8.0 & -6.6$\pm$0.2 & -6.2$\pm$0.1 & - & -7.7$\pm$0.2\\
SDSSJ0806+3055${}^{(z,\overline{q})}$ & He & 7017.0 & 0.6 & 8.0 & -7.0$\pm$0.3 & -6.8$\pm$0.2 & - & -7.9$\pm$0.3\\
SDSSJ0117+0021${}^{(z,\overline{q})}$ & He & 6994.0 & 0.608 & 8.052 & -7.6$\pm$0.1 & -7.2$\pm$0.1 & - & -8.8$\pm$0.1\\
SDSSJ2238+0213${}^{(z,\overline{q})}$ & He & 6986.0 & 0.6 & 8.0 & -7.5$\pm$0.2 & -7.4$\pm$0.2 & - & -8.6$\pm$0.2\\
SDSSJ0447+1124${}^{(z,\overline{q})}$ & He & 6966.0 & 0.492 & 7.858 & -7.7$\pm$0.3 & -7.3$\pm$0.3 & - & -9.0$\pm$0.3\\
SDSSJ1443+3014${}^{(z,\overline{q})}$ & He & 6955.0 & 0.6 & 8.0 & -7.1$\pm$0.2 & -6.5$\pm$0.2 & - & -8.1$\pm$0.3\\
SDSSJ0150+1354${}^{(z,\overline{q})}$ & He & 6953.0 & 0.741 & 8.262 & -6.8$\pm$0.2 & -6.0$\pm$0.1 & - & -7.7$\pm$0.2\\
SDSSJ0010-0430${}^{(z,\overline{q})}$ & He & 6903.0 & 0.651 & 8.122 & -7.0$\pm$0.1 & -6.7$\pm$0.1 & - & -8.5$\pm$0.1\\
SDSSJ0818+1247${}^{(z,\overline{q})}$ & He & 6895.0 & 0.6 & 8.0 & -7.8$\pm$0.4 & -7.2$\pm$0.3 & - & -9.0$\pm$0.3\\
SDSSJ1112+0700${}^{(z,\overline{q})}$ & He & 6891.0 & 0.497 & 7.867 & -8.5$\pm$0.5 & -7.6$\pm$0.3 & - & -9.7$\pm$0.2\\
SDSSJ1554+1735${}^{(z,\overline{q})}$ & He & 6847.0 & 0.721 & 8.231 & -7.6$\pm$0.1 & -7.1$\pm$0.1 & - & -8.4$\pm$0.1\\
SDSSJ0806+4058${}^{(z,\overline{q})}$ & He & 6808.0 & 0.6 & 8.0 & -7.49$\pm$0.08 & -7.38$\pm$0.12 & - & -8.49$\pm$0.08\\
SDSSJ1134+1542${}^{(z,\overline{q})}$ & He & 6806.0 & 0.6 & 8.0 & -7.3$\pm$0.4 & -6.9$\pm$0.1 & - & -8.5$\pm$0.1\\
SDSSJ1549+2633${}^{(z,\overline{q})}$ & He & 6794.0 & 0.614 & 8.063 & -8.0$\pm$0.2 & -7.9$\pm$0.1 & - & -9.6$\pm$0.2\\
SDSSJ0252-0401${}^{(z,\overline{q})}$ & He & 6773.0 & 0.53 & 7.924 & -8.0$\pm$0.2 & -6.8$\pm$0.1 & - & -8.9$\pm$0.1\\
SDSSJ1017+2419${}^{(z,\overline{q})}$ & He & 6772.0 & 0.84 & 8.409 & -7.3$\pm$0.2 & -6.7$\pm$0.2 & - & -8.4$\pm$0.2\\
SDSSJ0148-0112${}^{(z,\overline{q})}$ & He & 6760.0 & 0.6 & 8.0 & -7.9$\pm$0.5 & -7.3$\pm$0.3 & - & -9.0$\pm$0.2\\
SDSSJ0144+0305${}^{(z,\overline{q})}$ & He & 6753.0 & 0.6 & 8.0 & -7.2$\pm$0.3 & -7.1$\pm$0.3 & - & -8.3$\pm$0.3\\
SDSSJ0908+4119${}^{(z,\overline{q})}$ & He & 6746.0 & 0.6 & 8.0 & -7.1$\pm$0.3 & -6.8$\pm$0.2 & - & -8.7$\pm$0.3\\
SDSSJ1626+3303${}^{(z,\overline{q})}$ & He & 6715.0 & 0.6 & 8.0 & -7.5$\pm$0.3 & -7.1$\pm$0.1 & - & -8.6$\pm$0.2\\
SDSSJ1329+1301${}^{(z,\overline{q})}$ & He & 6706.0 & 0.381 & 7.636 & -8.0$\pm$0.3 & -7.4$\pm$0.1 & - & -8.9$\pm$0.2\\
SDSSJ1158+1845${}^{(z,\overline{q})}$ & He & 6696.0 & 0.6 & 8.04 & -7.4$\pm$0.2 & -6.4$\pm$0.1 & - & -8.6$\pm$0.2\\
SDSSJ1220+0929${}^{(z,\overline{q})}$ & He & 6677.0 & 0.6 & 8.0 & -7.4$\pm$0.1 & -6.8$\pm$0.1 & - & -8.4$\pm$0.1\\
SDSSJ0929+4247${}^{(z,\overline{q})}$ & He & 6676.0 & 0.489 & 7.853 & -7.1$\pm$0.2 & -6.5$\pm$0.2 & - & -8.3$\pm$0.2\\
SDSSJ0937+5228${}^{(z,\overline{q})}$ & He & 6660.0 & 0.585 & 8.015 & -7.4$\pm$0.2 & -6.9$\pm$0.1 & - & -8.4$\pm$0.1\\
SDSSJ1624+3310${}^{(z,\overline{q})}$ & He & 6654.0 & 0.6 & 8.0 & -8.0$\pm$0.4 & -7.2$\pm$0.1 & - & -9.2$\pm$0.3\\
SDSSJ1500+2315${}^{(z,\overline{q})}$ & He & 6630.0 & 0.6 & 8.0 & -7.0$\pm$0.4 & -6.6$\pm$0.2 & - & -8.2$\pm$0.4\\
SDSSJ0843+5614${}^{(z,\overline{q})}$ & He & 6624.0 & 0.761 & 8.292 & -7.5$\pm$0.2 & -6.9$\pm$0.2 & - & -8.7$\pm$0.2\\
SDSSJ0744+4408${}^{(z,\overline{q})}$ & He & 6612.0 & 0.6 & 8.0 & -7.5$\pm$0.2 & -7.4$\pm$0.2 & - & -8.6$\pm$0.2\\
SDSSJ1211+2326${}^{(z,\overline{q})}$ & He & 6609.0 & 1.005 & 8.663 & -6.9$\pm$0.2 & -6.7$\pm$0.1 & - & -8.3$\pm$0.2\\
SDSSJ1157+6138${}^{(z,\overline{q})}$ & He & 6607.0 & 0.6 & 8.0 & -8.2$\pm$0.5 & -7.6$\pm$0.5 & - & -9.3$\pm$0.3\\
SDSSJ0234-0510${}^{(z,\overline{q})}$ & He & 6601.0 & 0.6 & 8.0 & -7.3$\pm$0.2 & -6.9$\pm$0.2 & - & -8.5$\pm$0.2\\
SDSSJ1546+3009${}^{(z,\overline{q})}$ & He & 6600.0 & 0.6 & 8.0 & -7.19$\pm$0.12 & -7.19$\pm$0.15 & - & -8.4$\pm$0.12\\
SDSSJ1428+4403${}^{(z,\overline{q})}$ & He & 6574.0 & 0.614 & 8.064 & -8.9$\pm$0.2 & -7.9$\pm$0.1 & - & -9.5$\pm$0.1\\
SDSSJ0906+1141${}^{(z,\overline{q})}$ & He & 6556.0 & 0.6 & 8.0 & -7.3$\pm$0.2 & -6.9$\pm$0.2 & - & -8.5$\pm$0.2\\
SDSSJ1610+4006${}^{(z,\overline{q})}$ & He & 6552.0 & 0.6 & 8.0 & -7.2$\pm$0.2 & -7.3$\pm$0.1 & - & -8.7$\pm$0.3\\
SDSSJ0053+3115${}^{(z,\overline{q})}$ & He & 6548.0 & 0.6 & 8.0 & -7.6$\pm$0.5 & -7.6$\pm$0.5 & - & -8.8$\pm$0.5\\
SDSSJ1347+1415${}^{(z,\overline{q})}$ & He & 6520.0 & 0.696 & 8.194 & -7.3$\pm$0.2 & -7.1$\pm$0.1 & - & -8.6$\pm$0.2\\
SDSSJ1421+1843${}^{(z,\overline{q})}$ & He & 6517.0 & 0.562 & 7.978 & -7.3$\pm$0.2 & -6.7$\pm$0.1 & - & -8.5$\pm$0.1\\
SDSSJ2333+1058${}^{(z,\overline{q})}$ & He & 6515.0 & 0.6 & 8.0 & -7.6$\pm$0.4 & -7.0$\pm$0.2 & - & -8.7$\pm$0.3\\
SDSSJ2235-0056${}^{(z,\overline{q})}$ & He & 6514.0 & 0.758 & 8.288 & -7.2$\pm$0.2 & -7.1$\pm$0.1 & - & -8.4$\pm$0.2\\
SDSSJ1540+5352${}^{(z,\overline{q})}$ & He & 6500.0 & 0.6 & 8.0 & -7.5$\pm$0.3 & -6.8$\pm$0.1 & - & -8.7$\pm$0.3\\
SDSSJ1218+0023${}^{(z,\overline{q})}$ & He & 6500.0 & 0.776 & 8.314 & -8.5$\pm$0.3 & -7.8$\pm$0.2 & - & -9.2$\pm$0.2\\
SDSSJ1616+3303${}^{(z,\overline{q})}$ & He & 6491.0 & 0.569 & 7.991 & -7.1$\pm$0.2 & -6.7$\pm$0.1 & - & -8.2$\pm$0.1\\
SDSSJ1303+4055${}^{(z,\overline{q})}$ & He & 6481.0 & 0.618 & 8.071 & -7.9$\pm$0.3 & -7.2$\pm$0.2 & - & -8.7$\pm$0.3\\
SDSSJ0002+3209${}^{(z,\overline{q})}$ & He & 6466.0 & 0.737 & 8.257 & -7.9$\pm$0.2 & -7.4$\pm$0.2 & - & -9.1$\pm$0.2\\
SDSSJ1339+2643${}^{(z,\overline{q})}$ & He & 6452.0 & 0.622 & 8.077 & -8.4$\pm$0.2 & -7.6$\pm$0.1 & - & -9.0$\pm$0.2\\
SDSSJ0830-0319${}^{(z,\overline{q})}$ & He & 6424.0 & 0.428 & 7.736 & -8.2$\pm$0.1 & -8.2$\pm$0.1 & - & -9.4$\pm$0.1\\
SDSSJ1604+1830${}^{(z,\overline{q})}$ & He & 6421.0 & 0.636 & 8.1 & -8.5$\pm$0.3 & -8.4$\pm$0.2 & - & -9.6$\pm$0.1\\
SDSSJ1254+3551${}^{(z,\overline{q})}$ & He & 6417.0 & 0.597 & 8.036 & -8.1$\pm$0.2 & -7.6$\pm$0.2 & - & -9.7$\pm$0.2\\
SDSSJ1038+0432${}^{(z,\overline{q})}$ & He & 6363.0 & 0.6 & 8.0 & -7.0$\pm$0.2 & -6.6$\pm$0.2 & - & -7.6$\pm$0.3\\
SDSSJ1430-0151${}^{(z,\overline{q})}$ & He & 6344.0 & 0.621 & 8.076 & -6.4$\pm$0.1 & -6.0$\pm$0.1 & - & -7.6$\pm$0.1\\
SDSSJ2330+2805${}^{(z,\overline{q})}$ & He & 6344.0 & 0.6 & 8.0 & -8.1$\pm$0.4 & -7.5$\pm$0.2 & - & -9.3$\pm$0.2\\
SDSSJ1024+4531${}^{(z,\overline{q})}$ & He & 6339.0 & 0.803 & 8.356 & -7.6$\pm$0.2 & -7.3$\pm$0.2 & - & -8.4$\pm$0.2\\
SDSSJ0933+6334${}^{(z,\overline{q})}$ & He & 6337.0 & 0.6 & 8.0 & -7.0$\pm$0.3 & -7.0$\pm$0.3 & - & -8.6$\pm$0.3\\
SDSSJ1448+1047${}^{(z,\overline{q})}$ & He & 6331.0 & 0.682 & 8.173 & -7.8$\pm$0.2 & -7.5$\pm$0.1 & - & -9.1$\pm$0.2\\
SDSSJ0939+4136${}^{(z,\overline{q})}$ & He & 6321.0 & 0.848 & 8.422 & -6.6$\pm$0.2 & -6.8$\pm$0.2 & - & -8.1$\pm$0.2\\
SDSSJ1102+2827${}^{(z,\overline{q})}$ & He & 6320.0 & 0.6 & 8.0 & -6.7$\pm$0.4 & -6.2$\pm$0.2 & - & -7.8$\pm$0.4\\
SDSSJ1230+3143${}^{(z,\overline{q})}$ & He & 6310.0 & 0.398 & 7.676 & -8.3$\pm$0.3 & -8.1$\pm$0.3 & - & -9.6$\pm$0.3\\
SDSSJ0047+1628${}^{(z,\overline{q})}$ & He & 6300.0 & 0.6 & 8.0 & -6.7$\pm$0.2 & -6.8$\pm$0.2 & - & -8.0$\pm$0.2\\
SDSSJ0851+1543${}^{(z,\overline{q})}$ & He & 6284.0 & 0.649 & 8.12 & -8.2$\pm$0.2 & -7.4$\pm$0.2 & - & -8.6$\pm$0.1\\
SDSSJ1404+3620${}^{(z,\overline{q})}$ & He & 6284.0 & 0.709 & 8.216 & -8.3$\pm$0.2 & -7.3$\pm$0.1 & - & -8.7$\pm$0.1\\
SDSSJ0948+3008${}^{(z,\overline{q})}$ & He & 6284.0 & 0.228 & 7.21 & -8.1$\pm$0.2 & -7.9$\pm$0.2 & - & -9.1$\pm$0.2\\
SDSSJ1014+2827${}^{(z,\overline{q})}$ & He & 6269.0 & 0.6 & 8.0 & -6.5$\pm$0.3 & -6.4$\pm$0.2 & - & -7.9$\pm$0.3\\
SDSSJ1257-0310${}^{(z,\overline{q})}$ & He & 6269.0 & 0.6 & 8.0 & -7.4$\pm$0.2 & -7.1$\pm$0.1 & - & -8.6$\pm$0.1\\
SDSSJ0201+2015${}^{(z,\overline{q})}$ & He & 6250.0 & 0.63 & 8.091 & -8.0$\pm$0.2 & -7.3$\pm$0.2 & - & -9.0$\pm$0.2\\
SDSSJ0108-0537${}^{(z,\overline{q})}$ & He & 6250.0 & 0.241 & 7.256 & -7.9$\pm$0.3 & -8.2$\pm$0.3 & - & -8.5$\pm$0.3\\
SDSSJ1549+1906${}^{(z,\overline{q})}$ & He & 6246.0 & 0.6 & 8.0 & -7.3$\pm$0.3 & -7.0$\pm$0.1 & - & -8.7$\pm$0.2\\
SDSSJ0116+2050${}^{(z,\overline{q})}$ & He & 6245.0 & 0.6 & 8.0 & -7.6$\pm$0.1 & -7.4$\pm$0.1 & - & -8.8$\pm$0.1\\
SDSSJ2238-0113${}^{(z,\overline{q})}$ & He & 6228.0 & 0.6 & 8.0 & -8.0$\pm$0.2 & -7.6$\pm$0.2 & - & -9.4$\pm$0.2\\
SDSSJ1019+3535${}^{(z,\overline{q})}$ & He & 6224.0 & 0.6 & 8.0 & -7.9$\pm$0.4 & -7.0$\pm$0.4 & - & -8.8$\pm$0.4\\
SDSSJ1158+5448${}^{(z,\overline{q})}$ & He & 6213.0 & 0.6 & 8.0 & -7.8$\pm$0.4 & -7.3$\pm$0.2 & - & -9.0$\pm$0.3\\
SDSSJ0114+3505${}^{(z,\overline{q})}$ & He & 6209.0 & 0.6 & 8.0 & -7.1$\pm$0.1 & -7.0$\pm$0.1 & - & -8.4$\pm$0.1\\
SDSSJ1543+2024${}^{(z,\overline{q})}$ & He & 6206.0 & 0.6 & 8.0 & -7.6$\pm$0.4 & -7.1$\pm$0.2 & - & -8.7$\pm$0.2\\
SDSSJ1524+4049${}^{(z,\overline{q})}$ & He & 6203.0 & 0.701 & 8.203 & -7.5$\pm$0.2 & -7.1$\pm$0.1 & - & -8.4$\pm$0.1\\
SDSSJ1314+3748${}^{(z,\overline{q})}$ & He & 6201.0 & 0.819 & 8.379 & -7.1$\pm$0.4 & -6.9$\pm$0.2 & - & -8.2$\pm$0.4\\
SDSSJ1534+1242${}^{(z,\overline{q})}$ & He & 6197.0 & 0.6 & 8.0 & -6.7$\pm$0.3 & -6.9$\pm$0.2 & - & -7.9$\pm$0.3\\
SDSSJ1356+2416${}^{(z,\overline{q})}$ & He & 6173.0 & 0.58 & 8.01 & -8.2$\pm$0.2 & -7.3$\pm$0.2 & - & -9.1$\pm$0.2\\
SDSSJ1336+3547${}^{(z,\overline{q})}$ & He & 6172.0 & 0.609 & 8.057 & -7.9$\pm$0.1 & -7.1$\pm$0.1 & - & -8.9$\pm$0.1\\
SDSSJ1005+2244${}^{(z,\overline{q})}$ & He & 6165.0 & 0.6 & 8.0 & -7.8$\pm$0.5 & -7.9$\pm$0.5 & - & -9.1$\pm$0.4\\
SDSSJ2340+0817${}^{(z,\overline{q})}$ & He & 6151.0 & 0.661 & 8.141 & -7.2$\pm$0.3 & -7.4$\pm$0.3 & - & -8.7$\pm$0.2\\
SDSSJ1033+1809${}^{(z,\overline{q})}$ & He & 6147.0 & 0.743 & 8.267 & -7.6$\pm$0.3 & -7.7$\pm$0.3 & - & -8.6$\pm$0.3\\
SDSSJ2109-0039${}^{(z,\overline{q})}$ & He & 6132.0 & 0.572 & 7.997 & -7.6$\pm$0.4 & -7.7$\pm$0.4 & - & -8.8$\pm$0.5\\
SDSSJ1542+4650${}^{(z,\overline{q})}$ & He & 6130.0 & 0.744 & 8.269 & -6.9$\pm$0.1 & -6.5$\pm$0.1 & - & -8.1$\pm$0.2\\
SDSSJ2357+2348${}^{(z,\overline{q})}$ & He & 6117.0 & 0.6 & 8.0 & -7.8$\pm$0.3 & -7.3$\pm$0.1 & - & -9.2$\pm$0.1\\
SDSSJ0158-0942${}^{(z,\overline{q})}$ & He & 6115.0 & 0.567 & 7.99 & -8.8$\pm$0.5 & -8.7$\pm$0.5 & - & -9.9$\pm$0.2\\
SDSSJ1024+1014${}^{(z,\overline{q})}$ & He & 6105.0 & 0.6 & 8.0 & -6.6$\pm$0.3 & -6.1$\pm$0.3 & - & -7.7$\pm$0.3\\
SDSSJ0044+0418${}^{(z,\overline{q})}$ & He & 6104.0 & 0.711 & 8.22 & -8.4$\pm$0.2 & -8.3$\pm$0.3 & - & -9.8$\pm$0.1\\
SDSSJ1330+3029${}^{(z,\overline{q})}$ & He & 6100.0 & 0.6 & 8.0 & -7.3$\pm$0.06 & -7.15$\pm$0.1 & - & -8.4$\pm$0.06\\
SDSSJ0013+1109${}^{(z,\overline{q})}$ & He & 6090.0 & 0.6 & 8.0 & -8.0$\pm$0.5 & -7.5$\pm$0.4 & - & -9.2$\pm$0.4\\
SDSSJ1017+3447${}^{(z,\overline{q})}$ & He & 6089.0 & 0.576 & 8.004 & -8.6$\pm$0.5 & -8.6$\pm$0.5 & - & -9.8$\pm$0.5\\
SDSSJ0208-0542${}^{(z,\overline{q})}$ & He & 6085.0 & 0.6 & 8.0 & -7.1$\pm$0.3 & -7.2$\pm$0.3 & - & -8.2$\pm$0.3\\
SDSSJ1245+0822${}^{(z,\overline{q})}$ & He & 6074.0 & 0.6 & 8.0 & -7.6$\pm$0.2 & -7.0$\pm$0.2 & - & -8.3$\pm$0.2\\
SDSSJ2340+0124${}^{(z,\overline{q})}$ & He & 6072.0 & 0.574 & 8.0 & -8.0$\pm$0.2 & -7.2$\pm$0.1 & - & -8.8$\pm$0.1\\
SDSSJ1043+3516${}^{(z,\overline{q})}$ & He & 6069.0 & 0.657 & 8.135 & -7.6$\pm$0.1 & -7.5$\pm$0.1 & - & -9.3$\pm$0.1\\
SDSSJ1545+5236${}^{(z,\overline{q})}$ & He & 6068.0 & 0.639 & 8.107 & -8.0$\pm$0.2 & -7.5$\pm$0.1 & - & -8.9$\pm$0.1\\
SDSSJ2352+1922${}^{(z,\overline{q})}$ & He & 6067.0 & 0.6 & 8.0 & -7.2$\pm$0.3 & -6.8$\pm$0.1 & - & -8.3$\pm$0.2\\
SDSSJ1132+3323${}^{(z,\overline{q})}$ & He & 6062.0 & 0.6 & 8.0 & -7.1$\pm$0.2 & -6.8$\pm$0.2 & - & -8.3$\pm$0.2\\
SDSSJ0925+3130${}^{(z,\overline{q})}$ & He & 6050.0 & 0.6 & 8.0 & -7.6$\pm$0.2 & -7.4$\pm$0.2 & - & -8.6$\pm$0.2\\
SDSSJ0800+2242${}^{(z,\overline{q})}$ & He & 6049.0 & 0.6 & 8.0 & -8.7$\pm$0.5 & -8.7$\pm$0.5 & - & -9.9$\pm$0.3\\
SDSSJ0838+2322${}^{(z,\overline{q})}$ & He & 6048.0 & 0.727 & 8.244 & -8.7$\pm$0.5 & -8.6$\pm$0.5 & - & -9.8$\pm$0.2\\
SDSSJ1158+5942${}^{(z,\overline{q})}$ & He & 6046.0 & 0.587 & 8.022 & -8.4$\pm$0.3 & -7.7$\pm$0.2 & - & -8.9$\pm$0.3\\
SDSSJ2157+1206${}^{(z,\overline{q})}$ & He & 6042.0 & 0.531 & 7.93 & -8.1$\pm$0.1 & -7.6$\pm$0.1 & - & -9.1$\pm$0.1\\
SDSSJ0939+5019${}^{(z,\overline{q})}$ & He & 6030.0 & 0.536 & 7.939 & -7.0$\pm$0.2 & -6.7$\pm$0.2 & - & -8.4$\pm$0.2\\
SDSSJ2225+2338${}^{(z,\overline{q})}$ & He & 6029.0 & 0.535 & 7.937 & -8.4$\pm$0.1 & -7.7$\pm$0.1 & - & -9.3$\pm$0.1\\
SDSSJ0144+1920${}^{(z,\overline{q})}$ & He & 6024.0 & 0.5 & 7.877 & -7.6$\pm$0.2 & -7.3$\pm$0.2 & - & -8.7$\pm$0.2\\
SDSSJ1040+2407${}^{(z,\overline{q})}$ & He & 6023.0 & 0.686 & 8.181 & -7.2$\pm$0.2 & -6.8$\pm$0.1 & - & -8.1$\pm$0.2\\
SDSSJ0721+3928${}^{(z,\overline{q})}$ & He & 6022.0 & 0.296 & 7.433 & -8.4$\pm$0.2 & -7.7$\pm$0.2 & - & -9.3$\pm$0.2\\
SDSSJ1229+0743${}^{(z,\overline{q})}$ & He & 6014.0 & 0.204 & 7.133 & -7.3$\pm$0.2 & -7.3$\pm$0.2 & - & -8.9$\pm$0.2\\
SDSSJ0135+1302${}^{(z,\overline{q})}$ & He & 6013.0 & 0.633 & 8.097 & -8.1$\pm$0.3 & -7.9$\pm$0.2 & - & -9.2$\pm$0.1\\
SDSSJ1217+1157${}^{(z,\overline{q})}$ & He & 6012.0 & 0.285 & 7.401 & -8.7$\pm$0.3 & -7.8$\pm$0.2 & - & -9.6$\pm$0.2\\
SDSSJ0913+4127${}^{(z,\overline{q})}$ & He & 6010.0 & 0.6 & 8.0 & -7.4$\pm$0.5 & -7.3$\pm$0.3 & - & -8.6$\pm$0.3\\
SDSSJ1308+0258${}^{(z,\overline{q})}$ & He & 6003.0 & 0.6 & 8.0 & -7.8$\pm$0.2 & -7.9$\pm$0.2 & - & -9.3$\pm$0.2\\
SDSSJ1205+3536${}^{(z,\overline{q})}$ & He & 6000.0 & 0.684 & 8.178 & -7.6$\pm$0.1 & -7.2$\pm$0.1 & - & -8.9$\pm$0.1\\
SDSSJ1345+1153${}^{(z,\overline{q})}$ & He & 6000.0 & 0.689 & 8.19 & -6.9$\pm$0.2 & -7.0$\pm$0.1 & - & -8.1$\pm$0.1\\
SDSSJ1627+4646${}^{(z,\overline{q})}$ & He & 6000.0 & 0.45 & 7.785 & -8.3$\pm$0.5 & -7.9$\pm$0.3 & - & -9.4$\pm$0.5\\
SDSSJ0741+3146${}^{(z,\overline{q})}$ & He & 5974.0 & 0.701 & 8.205 & -7.5$\pm$0.2 & -7.6$\pm$0.2 & - & -9.0$\pm$0.2\\
SDSSJ0842+1536${}^{(z,\overline{q})}$ & He & 5966.0 & 0.823 & 8.385 & -8.6$\pm$0.2 & -8.5$\pm$0.2 & - & -9.8$\pm$0.4\\
SDSSJ1535+1247${}^{(z,\overline{q})}$ & He & 5950.0 & 0.682 & 8.175 & -7.5$\pm$0.1 & -7.0$\pm$0.1 & - & -8.6$\pm$0.1\\
SDSSJ1401+3659${}^{(z,\overline{q})}$ & He & 5931.0 & 0.557 & 7.973 & -9.2$\pm$0.3 & -8.1$\pm$0.3 & - & -10.1$\pm$0.1\\
SDSSJ0823+0546${}^{(gg,\overline{aa})}$ & He & 5920.0 & 0.6 & 7.945 & -7.35$\pm$0.1 & -7.85$\pm$0.1 & - & -9.8$\pm$0.1\\
SDSSJ1405+2542${}^{(z,\overline{q})}$ & He & 5890.0 & 0.6 & 8.0 & -8.4$\pm$0.5 & -7.8$\pm$0.3 & - & -9.5$\pm$0.4\\
SDSSJ0004+0819${}^{(z,\overline{q})}$ & He & 5843.0 & 0.6 & 8.0 & -7.6$\pm$0.4 & -7.2$\pm$0.2 & - & -8.8$\pm$0.2\\
SDSSJ1105+0228${}^{(z,\overline{q})}$ & He & 5842.0 & 0.6 & 8.0 & -8.0$\pm$0.4 & -7.6$\pm$0.2 & - & -9.1$\pm$0.4\\
SDSSJ2110+0512${}^{(z,\overline{q})}$ & He & 5828.0 & 0.6 & 8.0 & -8.3$\pm$0.5 & -7.7$\pm$0.5 & - & -9.4$\pm$0.5\\
SDSSJ1706+2541${}^{(z,\overline{q})}$ & He & 5813.0 & 0.6 & 8.0 & -9.0$\pm$0.5 & -8.9$\pm$0.5 & - & -10.1$\pm$0.3\\
SDSSJ0019+2209${}^{(z,\overline{q})}$ & He & 5797.0 & 0.365 & 7.608 & -8.5$\pm$0.2 & -8.4$\pm$0.2 & - & -9.7$\pm$0.2\\
SDSSJ0006+0520${}^{(z,\overline{q})}$ & He & 5783.0 & 0.478 & 7.839 & -9.6$\pm$0.3 & -8.6$\pm$0.5 & - & -9.8$\pm$0.2\\
SDSSJ0908+5136${}^{(z,\overline{q})}$ & He & 5779.0 & 0.447 & 7.78 & -9.0$\pm$0.3 & -8.9$\pm$0.3 & - & -10.1$\pm$0.3\\
SDSSJ2343-0010${}^{(z,\overline{q})}$ & He & 5778.0 & 0.309 & 7.475 & -8.5$\pm$0.5 & -7.9$\pm$0.3 & - & -9.6$\pm$0.3\\
SDSSJ1103+4144${}^{(z,\overline{q})}$ & He & 5728.0 & 0.327 & 7.521 & -8.3$\pm$0.2 & -7.7$\pm$0.2 & - & -9.4$\pm$0.1\\
SDSSJ1102+0214${}^{(z,\overline{q})}$ & He & 5699.0 & 0.658 & 8.137 & -8.6$\pm$0.4 & -8.0$\pm$0.5 & - & -9.8$\pm$0.1\\
SDSSJ1259+4729${}^{(z,\overline{q})}$ & He & 5682.0 & 0.454 & 7.795 & -8.7$\pm$0.5 & -8.6$\pm$0.5 & - & -9.8$\pm$0.4\\
SDSSJ1535+1247${}^{(ff,\overline{cc})}$ & He & 5680.0 & 0.61 & 8.06 & -7.6$\pm$0.1 & -6.9$\pm$0.1 & -6.5$\pm$0.3 & -8.5$\pm$0.05\\
SDSSJ2231+0906${}^{(z,\overline{q})}$ & He & 5679.0 & 0.567 & 7.992 & -9.0$\pm$0.2 & -8.1$\pm$0.5 & - & -9.9$\pm$0.1\\
SDSSJ1259+3112${}^{(z,\overline{q})}$ & He & 5664.0 & 0.589 & 8.027 & -8.6$\pm$0.4 & -8.6$\pm$0.4 & - & -9.8$\pm$0.3\\
SDSSJ1351+2645${}^{(z,\overline{q})}$ & He & 5640.0 & 0.421 & 7.73 & -7.8$\pm$0.3 & -7.9$\pm$0.2 & - & -8.5$\pm$0.2\\
SDSSJ1429+3841${}^{(z,\overline{q})}$ & He & 5636.0 & 0.375 & 7.634 & -9.2$\pm$0.5 & -9.1$\pm$0.5 & - & -10.3$\pm$0.2\\
SDSSJ2230+1905${}^{(z,\overline{q})}$ & He & 5631.0 & 0.6 & 8.0 & -7.1$\pm$0.2 & -6.8$\pm$0.2 & - & -8.5$\pm$0.3\\
SDSSJ1055+3725${}^{(z,\overline{q})}$ & He & 5614.0 & 0.361 & 7.605 & -8.2$\pm$0.5 & -8.2$\pm$0.2 & - & -8.4$\pm$0.3\\
SDSSJ1006+1752${}^{(z,\overline{q})}$ & He & 5605.0 & 0.731 & 8.251 & -8.3$\pm$0.5 & -8.3$\pm$0.5 & - & -9.5$\pm$0.5\\
SDSSJ1321-0237${}^{(z,\overline{q})}$ & He & 5592.0 & 0.6 & 8.0 & -7.3$\pm$0.3 & -6.9$\pm$0.2 & - & -8.7$\pm$0.3\\
SDSSJ0126+2534${}^{(z,\overline{q})}$ & He & 5588.0 & 0.87 & 8.456 & -8.3$\pm$0.5 & -8.3$\pm$0.5 & - & -9.5$\pm$0.3\\
SDSSJ0758+1013${}^{(z,\overline{q})}$ & He & 5585.0 & 0.6 & 8.0 & -7.1$\pm$0.2 & -7.0$\pm$0.2 & - & -8.5$\pm$0.2\\
SDSSJ2328+0830${}^{(z,\overline{q})}$ & He & 5566.0 & 0.6 & 8.0 & -7.8$\pm$0.4 & -7.3$\pm$0.1 & - & -9.0$\pm$0.3\\
SDSSJ0512-0505${}^{(gg,\overline{aa})}$ & He & 5560.0 & 0.803 & 8.05 & -7.75$\pm$0.1 & -7.65$\pm$0.1 & - & -8.9$\pm$0.1\\
SDSSJ1502+3744${}^{(z,\overline{q})}$ & He & 5525.0 & 0.624 & 8.084 & -8.9$\pm$0.5 & -8.8$\pm$0.5 & - & -10.1$\pm$0.2\\
SDSSJ1342+1813${}^{(z,\overline{q})}$ & He & 5524.0 & 0.6 & 8.0 & -7.9$\pm$0.5 & -7.8$\pm$0.5 & - & -9.0$\pm$0.5\\
SDSSJ1537+3608${}^{(z,\overline{q})}$ & He & 5519.0 & 0.6 & 8.0 & -8.6$\pm$0.5 & -8.6$\pm$0.5 & - & -9.8$\pm$0.3\\
SDSSJ1019+2045${}^{(z,\overline{q})}$ & He & 5515.0 & 0.6 & 8.0 & -8.2$\pm$0.5 & -8.2$\pm$0.5 & - & -9.4$\pm$0.4\\
SDSSJ1411+3410${}^{(z,\overline{q})}$ & He & 5500.0 & 0.807 & 8.363 & -7.1$\pm$0.5 & -6.6$\pm$0.2 & - & -8.3$\pm$0.4\\
SDSSJ0924+4301${}^{(z,\overline{q})}$ & He & 5500.0 & 0.6 & 8.0 & -9.0$\pm$0.5 & -9.0$\pm$0.5 & - & -10.2$\pm$0.4\\
SDSSJ0916+2540${}^{(z,\overline{q})}$ & He & 5497.0 & 0.612 & 8.066 & -7.2$\pm$0.2 & -6.5$\pm$0.2 & - & -7.5$\pm$0.2\\
SDSSJ1032+1338${}^{(z,\overline{q})}$ & He & 5479.0 & 0.6 & 8.0 & -8.1$\pm$0.3 & -7.3$\pm$0.3 & - & -9.1$\pm$0.3\\
SDSSJ1641+1856${}^{(z,\overline{q})}$ & He & 5470.0 & 0.536 & 7.941 & -9.9$\pm$0.3 & -9.2$\pm$0.5 & - & -10.4$\pm$0.2\\
SDSSJ2123+0016${}^{(z,\overline{q})}$ & He & 5463.0 & 0.675 & 8.166 & -8.4$\pm$0.4 & -8.3$\pm$0.5 & - & -9.6$\pm$0.2\\
SDSSJ1238+2149${}^{(z,\overline{q})}$ & He & 5437.0 & 0.632 & 8.097 & -7.9$\pm$0.3 & -7.3$\pm$0.2 & - & -8.9$\pm$0.3\\
SDSSJ1257+3238${}^{(z,\overline{q})}$ & He & 5376.0 & 0.6 & 8.0 & -7.6$\pm$0.4 & -7.2$\pm$0.2 & - & -8.8$\pm$0.2\\
SDSSJ1158+0454${}^{(z,\overline{q})}$ & He & 5344.0 & 0.566 & 7.991 & -7.5$\pm$0.3 & -7.0$\pm$0.1 & - & -8.5$\pm$0.2\\
SDSSJ1144+1218${}^{(z,\overline{q})}$ & He & 5320.0 & 0.577 & 8.009 & -8.3$\pm$0.2 & -7.8$\pm$0.2 & - & -9.3$\pm$0.1\\
SDSSJ0052+1846${}^{(z,\overline{q})}$ & He & 5305.0 & 0.6 & 8.0 & -7.5$\pm$0.3 & -7.1$\pm$0.2 & - & -9.0$\pm$0.3\\
SDSSJ1649+2238${}^{(z,\overline{q})}$ & He & 5261.0 & 0.634 & 8.101 & -7.4$\pm$0.2 & -6.7$\pm$0.2 & - & -8.4$\pm$0.2\\
SDSSJ0913+2627${}^{(z,\overline{q})}$ & He & 5252.0 & 0.6 & 8.161 & -8.5$\pm$0.4 & -8.5$\pm$0.5 & - & -9.7$\pm$0.3\\
SDSSJ0739+3112${}^{(z,\overline{q})}$ & He & 5221.0 & 0.6 & 8.0 & -8.0$\pm$0.5 & -7.9$\pm$0.5 & - & -9.2$\pm$0.4\\
SDSSJ1518+0506${}^{(z,\overline{q})}$ & He & 5187.0 & 0.672 & 8.161 & -8.7$\pm$0.3 & -8.3$\pm$0.3 & - & -9.3$\pm$0.2\\
SDSSJ1046+1329${}^{(z,\overline{q})}$ & He & 5177.0 & 0.507 & 7.895 & -8.2$\pm$0.5 & -8.1$\pm$0.5 & - & -9.3$\pm$0.3\\
SDSSJ1350+1058${}^{(z,\overline{q})}$ & He & 5176.0 & 0.902 & 8.505 & -8.3$\pm$0.5 & -8.3$\pm$0.5 & - & -9.5$\pm$0.3\\
SDSSJ0807+4930${}^{(z,\overline{q})}$ & He & 5172.0 & 0.574 & 8.005 & -7.2$\pm$0.2 & -6.8$\pm$0.2 & - & -8.3$\pm$0.2\\
SDSSJ1316+1918${}^{(z,\overline{q})}$ & He & 5160.0 & 0.375 & 7.639 & -9.2$\pm$0.5 & -9.1$\pm$0.5 & - & -10.3$\pm$0.4\\
SDSSJ2304+2415${}^{(z,\overline{q})}$ & He & 5102.0 & 0.53 & 7.935 & -8.9$\pm$0.2 & -7.3$\pm$0.2 & - & -9.2$\pm$0.2\\
SDSSJ0056+2453${}^{(z,\overline{q})}$ & He & 5061.0 & 0.6 & 8.0 & -9.1$\pm$0.5 & -9.0$\pm$0.5 & - & -10.2$\pm$0.5\\
SDSSJ0736+4118${}^{(z,\overline{q})}$ & He & 5010.0 & 0.711 & 8.222 & -7.3$\pm$0.3 & -7.3$\pm$0.3 & - & -8.5$\pm$0.3\\
SDSSJ1147+5429${}^{(z,\overline{q})}$ & He & 5000.0 & 0.6 & 8.0 & -8.3$\pm$0.5 & -8.3$\pm$0.5 & - & -9.5$\pm$0.5\\
SDSSJ1224+2838${}^{(z,\overline{q})}$ & He & 4991.0 & 0.576 & 8.01 & -9.1$\pm$0.5 & -9.0$\pm$0.5 & - & -10.2$\pm$0.1\\
SDSSJ0744+4649${}^{(z,\overline{q})}$ & He & 4861.0 & 0.602 & 8.052 & -7.6$\pm$0.2 & -7.6$\pm$0.2 & - & -8.3$\pm$0.1\\
SDSSJ0852+3402${}^{(z,\overline{q})}$ & He & 4806.0 & 0.425 & 7.743 & -8.9$\pm$0.5 & -8.9$\pm$0.5 & - & -10.1$\pm$0.3\\
SDSSJ1152+5101${}^{(z,\overline{q})}$ & He & 4790.0 & 0.6 & 8.0 & -9.2$\pm$0.5 & -9.1$\pm$0.5 & - & -10.4$\pm$0.3\\
LHS2534${}^{(gg,\overline{cc})}$ & He & 4780.0 & 0.55 & 7.97 & -9.06$\pm$0.08 & -8.62$\pm$0.06 & - & -10.08$\pm$0.11\\
SDSSJ0744+1640${}^{(z,\overline{q})}$ & He & 4703.0 & 0.515 & 7.91 & -9.1$\pm$0.5 & -9.1$\pm$0.5 & - & -10.3$\pm$0.3\\
SDSSJ1226+2936${}^{(z,\overline{q})}$ & He & 4680.0 & 0.6 & 8.0 & -8.93$\pm$0.33 & -8.63$\pm$0.34 & - & -10.04$\pm$0.33\\
SDSSJ1636+1619${}^{(z,\overline{q})}$ & He & 4410.0 & 0.669 & 8.096 & -8.3$\pm$1.0 & -8.3$\pm$1.0 & - & -9.5$\pm$0.1\\
\hline
\end{longtable}
\begin{tablenotes}
\item References stellar properties:${}^{(a)}$\citet{Zuckerman2007} ${}^{(b)}$\citet{Klein2011} ${}^{(c)}$\citet{Zuckerman2011} ${}^{(d)}$\citet{Vennes2011b} ${}^{(e)}$\citet{Gaensicke2012} ${}^{(f)}$\citet{Klein2011} ${}^{(g)}$\citet{Dufour2012} ${}^{(h)}$\citet{Kawka2012} ${}^{(i)}$\citet{Vennes2013} ${}^{(j)}$\citet{Jura2012} ${}^{(k)}$\citet{Farihi2013} ${}^{(l)}$\citet{Xu2013} ${}^{(m)}$\citet{Xu2014b} ${}^{(n)}$\citet{Wilson2015} ${}^{(o)}$\citet{Raddi2015} ${}^{(p)}$\citet{Kawka2016} ${}^{(q)}$\citet{Hollands2017} ${}^{(r)}$\citet{Xu2017} ${}^{(s)}$\citet{Farihi2016b} ${}^{(t)}$\citet{Melis2017} ${}^{(u)}$\citet{Xu2019} ${}^{(v)}$\citet{Swan2019b} ${}^{(w)}$\citet{Fortin-Archambault2020} ${}^{(x)}$\citet{Izquierdo2021} ${}^{(y)}$\citet{Hoskin2020} ${}^{(z)}$\citet{GonzalezEgea2021} ${}^{(aa)}$\citet{Harrison2021b} ${}^{(bb)}$\citet{Klein2021} ${}^{(cc)}$\citet{Hollands2021} ${}^{(dd)}$\citet{Rogers2022} 
\item References abundances:${}^{(\overline{a})}$\citet{Farihi2010} ${}^{(\overline{b})}$\citet{Gianninas2011} ${}^{(\overline{c})}$\citet{Farihi2011} ${}^{(\overline{d})}$\citet{Xu2019} ${}^{(\overline{e})}$\citet{Kilic2020} ${}^{(\overline{f})}$\citet{Koester2014} ${}^{(\overline{g})}$\citet{Klein2011} ${}^{(\overline{i})}$\citet{Dufour2012} ${}^{(\overline{j})}$\citet{Tremblay2011} ${}^{(\overline{k})}$\citet{Fortin-Archambault2020} ${}^{(\overline{l})}$\citet{Kleinman2013} ${}^{(\overline{m})}$\citet{Coutu2019} ${}^{(\overline{n})}$\citet{Bergeron2011} ${}^{(\overline{o})}$\citet{Farihi2016b} ${}^{(\overline{p})}$\citet{Jura2012} ${}^{(\overline{q})}$\citet{Xu2014b} ${}^{(\overline{r})}$\citet{Leggett2018} ${}^{(\overline{s})}$\citet{Becklin2005} ${}^{(\overline{t})}$\citet{Kawka2012} ${}^{(\overline{u})}$\citet{Vennes2013} ${}^{(\overline{v})}$\citet{Wilson2015} ${}^{(\overline{w})}$\citet{Raddi2015} ${}^{(\overline{x})}$\citet{Blouin2019} ${}^{(\overline{y})}$\citet{Kawka2016} ${}^{(\overline{z})}$\citet{Hollands2017} ${}^{(\overline{aa})}$\citet{Swan2019b} ${}^{(\overline{bb})}$\citet{Izquierdo2021} ${}^{(\overline{cc})}$\citet{Hoskin2020} ${}^{(\overline{dd})}$\citet{GonzalezEgea2021} ${}^{(\overline{ee})}$\citet{Klein2021} ${}^{(\overline{ff})}$\citet{Hollands2021} ${}^{(\overline{gg})}$\citet{Harrison2021b} ${}^{(\overline{hh})}$\citet{Rogers2022} 
\end{tablenotes}
\end{center}
\twocolumn

\end{appendix}

\end{document}